\documentclass[iop, revtex4]{emulateapj}
\slugcomment{Accepted to {\it The Astrophysical Journal}}
\shortauthors{Martin et al.}
\shorttitle{Brown Dwarf Surface Gravities}

\usepackage{color,amsmath, pdflscape, perpage, float, rotating} 
\MakePerPage{footnote}

\newcommand{\kii}{\ion{K}{1}}
\newcommand{\nai}{\ion{Na}{1}}

\newcommand{\wat}{H$_2$O}

\newcommand{\kms}{km~s$^{-1}$}
\newcommand{\masyr}{mas~yr$^{-1}$}

\newcommand{\microm}{$\mu$m}

\newcommand{\mua}{$\mu_{\alpha}$}
\newcommand{\mud}{$\mu_{\delta}$}

\begin{document}

\title{Surface Gravities for 228 M, L, and T dwarfs in the NIRSPEC Brown Dwarf Spectroscopic Survey\textsuperscript{*}}

\author{
Emily C.\ Martin\altaffilmark{1}, Gregory N.\ Mace\altaffilmark{1,2}, Ian S.\ McLean\altaffilmark{1}, Sarah E.\ Logsdon\altaffilmark{1}, Emily L.\ Rice\altaffilmark{3,4,5}, J. Davy Kirkpatrick\altaffilmark{6}, Adam J. Burgasser\altaffilmark{7}, Mark R. McGovern\altaffilmark{8}, and Lisa Prato\altaffilmark{9}
}

\altaffiltext{1}{Department of Physics and Astronomy, University of California Los Angeles, 430 Portola Plaza, Box 951547, Los Angeles, CA, 90095-1547, USA; emartin@astro.ucla.edu}
\altaffiltext{2}{Department of Astronomy, UT Austin, 2515 Speedway, Stop C1400, Austin, TX 78712-1205}
\altaffiltext{3}{Department of Engineering Science \& Physics, College of Staten Island, 2800 Victory Blvd, Staten Island, NY 10301, USA}
\altaffiltext{4}{Department of Astrophysics, American Museum of Natural History, Central Park West at 79th St, New York, NY 10034}
\altaffiltext{5}{Physics Program, The Graduate Center, City University of New York, 365 Fifth Avenue, New York, NY 10016}
\altaffiltext{6}{Infrared Processing and Analysis Center, MS 100-22, California Institute of Technology, Pasadena, CA 91125, USA}
\altaffiltext{7}{Center for Astrophysics and Space Science, University of California San Diego, La Jolla, CA 92093, USA}
\altaffiltext{8}{Math \& Sciences Division, Antelope Valley College, 3041 West Ave K, Lancaster, CA 93536}
\altaffiltext{9}{Lowell Observatory, 1400 West Mars Hill Road, Flagstaff, AZ 86001}

\begin{abstract}

We combine 131 new medium-resolution (R $\sim$ 2000) $J$-band spectra of M, L, and T dwarfs from the Keck NIRSPEC Brown Dwarf Spectroscopic Survey (BDSS) with 97 previously published BDSS spectra to study surface-gravity-sensitive indices for 228 low-mass stars and brown dwarfs spanning spectral types M5--T9. Specifically, we use an established set of spectral indices to determine surface gravity classifications for all M6--L7 objects in our sample by measuring equivalent widths (EW) of the \kii \ lines at 1.1692, 1.1778, 1.2529 \microm , and the 1.2 \microm \ FeH$_J$ absorption index.  Our results are consistent with previous surface gravity measurements, showing a distinct double peak - at $\sim$ L5 and T5 - in \kii \ EW as a function of spectral type. We analyze \kii \ EWs of 73 objects of known ages and find a linear trend between log(Age) and EW. From this relationship, we assign age ranges to the very low gravity, intermediate gravity, and field gravity designations for spectral types M6--L0. Interestingly, the ages probed by these designations remain broad, change with spectral type, and depend on the gravity sensitive index used. Gravity designations are useful indicators of the possibility of youth, but current datasets cannot be used to provide a precise age estimate.

\end{abstract}

\footnotetext[*]{The data presented herein were obtained at the W.M. Keck Observatory, which is operated as a scientific partnership among the California Institute of Technology, the University of California and the National Aeronautics and Space Administration. The Observatory was made possible by the generous financial support of the W.M. Keck Foundation.}

\keywords{brown dwarfs --- infrared: stars --- stars: atmospheres --- stars: low-mass --- surveys} 
\section{Introduction}
\label{sec:intro}

Brown dwarfs are the lowest-mass products of star formation, with masses so low that they will never achieve stable hydrogen fusion in their cores \citep{kumar1962, kumar1963, hayashi1963}. Substellar objects are classified by their spectral morphology as types M, L, T, and Y, a sequence which represents both a decrease in effective temperature and changes in chemical abundances. Since their discovery 20 years ago \citep{nakajima1995, rebolo1995}, most brown dwarfs have been found through infrared large area surveys such as the Two Micron All Sky Survey (2MASS; \citealt{skrutskie2006}), the Sloan Digital Sky Survey (SDSS; \citealt{york2000}), the United Kingdom Infrared Deep Sky Survey (UKIDSS; \citealt{lawrence2007}), and the Wide-field Infrared Survey Explorer (WISE; \citealt{wright2010}), among others. See, for example, \citet{kirkpatrick1991, kirkpatrick1999, hawley2002, burgasser2006a, chiu2006, burningham2010, cushing2011} for details on brown dwarf discoveries made by the various surveys. 

Extensive follow-up using both optical and infrared imaging and spectroscopy has enabled astronomers to begin characterizing the physical properties of brown dwarfs, primarily through comparisons to atmospheric and evolutionary models, like those of \citet{burrows2001}, \citet{saumon2008}, \citet{allard2012}, and \citet{baraffe2015}. It is possible to constrain the effective temperatures, surface gravities, and metallicities of brown dwarfs within the limits of current models, e.g. \citet{cushing2008, rice2010}. As the number of confirmed brown dwarfs has increased, the properties typical of field brown dwarfs have been constrained, outliers have been recognized, and methods of identifying extremely young or old objects have emerged (see \citealt{kirkpatrick2010, allers2013} and references therein). 

\subsection{Surface Gravity as an Age Indicator}

Unlike stars, brown dwarfs contract and cool as they age, producing a degeneracy between the mass, age, and temperature such that temperature alone cannot reveal the mass or age of a given brown dwarf. For example, an L dwarf could be a young, planetary-mass brown dwarf, a moderately aged high-mass brown dwarf, or an old low-mass star. Brown dwarfs contract considerably in their first $\sim$ 300 Myr and significantly increase their surface gravity ($g=GM/R^2$) from log $g \sim 3.5$ to log $g \sim 5$ in units of cm~s$^{-2}$ \citep{burrows2001}. Obtaining a surface gravity estimate is an important step towards disentangling the mass and age of a brown dwarf.

Surface gravity affects several features in the optical and near infrared (NIR) spectra of brown dwarfs. Photospheric pressure, which is proportional to surface gravity assuming hydrostatic equilibrium, broadens atomic features and influences the chemical pathways of both atomic and molecular species \citep{lodders1999}. Neutral alkali lines such as \kii \  and \nai \ are weaker in low-gravity objects compared to higher gravity objects at similar spectral types because lower photospheric pressure decreases the column densities of the absorbing species above the photosphere, causing the absorption features to appear weaker in low gravity dwarfs. FeH absorption also appears weaker in lower gravity objects, while VO shows stronger absorption at lower gravity, as noted by \citet{mcgovern2004} and \citet{allers2013}. Additionally, the overall shape of the H-band spectral energy distribution is much ``peakier" at lower gravities \citep{lucas2001, allers2007, looper2008, rice2011}, likely due to lower H$_2$ collision induced absorption (CIA), which is a result of the lower photospheric pressure at lower gravities \citep{kirkpatrick2006, saumon2008}. 

\citet{kirkpatrick2005} proposed a scheme in which a gravity classification (i.e. $\alpha$, $\beta$, $\gamma$, $\delta$) is appended to the spectral type of a brown dwarf as a means of distinguishing between field, intermediate, low, and very-low gravity objects with similar temperatures. For each spectral type, the gravity sequence acts as a proxy for an age sequence, and low-gravity objects of a particular spectral type are younger than their field counterparts at the same spectral type. \citet{cruz2009} explored this gravity classification scheme using red-optical spectroscopy of 23 L dwarfs, primarily distinguishing the young objects from field-age objects by the weakness of their alkali lines, though also using the FeH, CrH, TiO and VO absorption bands as diagnostics. 

\citet{allers2013}, hereafter A13, were the first to present a systematic technique using NIR spectroscopy to determine surface gravities of low-mass stars and brown dwarfs. A13 defines spectral indices and pseudo-equivalent widths (EWs) of various gravity-sensitive features in lower resolution NIR spectra to classify the spectra into three groups: low (VL-G), intermediate (INT-G), and high (FLD-G) gravity objects, roughly corresponding to young ($\lesssim$ 30 Myr), intermediate ($\sim$30--200 Myr), and field age ($\gtrsim$ 200 Myr) objects. Because brown dwarfs are significantly brighter in the NIR than the optical, a NIR gravity classification scheme is applicable to more targets. A13 determined gravity classifications for 73 low-mass stars and brown dwarfs showing signs of youth. \citet{gagne2015b} applied the method prescribed in A13 to 182 objects of spectral types M4--L7 in the search for low-mass members of young moving groups. 

In this paper we follow up on prior NIR spectroscopy by our group and use a modified A13 method to determine surface gravities for 228 M, L, and T dwarfs.  Twenty of these targets overlap with the A13 sample, and 5 objects overlap with the \citet{gagne2015b} sample. Many previously unpublished NIR spectra from the NIRSPEC Brown Dwarf Spectroscopic Survey (BDSS) are reported and analyzed. 
 
\subsection{The NIRSPEC Brown Dwarf Spectroscopic Survey}

In 1999, the Near-Infrared Spectrometer (NIRSPEC; \citealt{mclean1998}) was commissioned for the W.M. Keck II 10-m telescope on Mauna Kea in Hawaii. NIRSPEC was built at the University of California, Los Angeles (UCLA), and designed for both medium (R=$\lambda / \Delta \lambda \sim 2000$) and high (R$\sim$20,000) resolution spectroscopy in the 1--5 \microm \  regime. The BDSS was one of the key science drivers for NIRSPEC. The primary goal of the BDSS as outlined in \citet{mclean2003} is to gather a large suite of NIR spectra of low-mass stars and brown dwarfs in order to examine their spectral properties and make comparisons to evolutionary and atmospheric models. The low temperatures (T $\lesssim$ 2500 K) of brown dwarfs make them excellent targets for NIR studies. Over the past 15 years, the BDSS team has gathered a large spectroscopic database of brown dwarfs and low-mass stars (see \S 2, Appendix), much of which is presented in \citet{mclean2001, mclean2003, mcgovern2004, mclean2007, rice2010, prato2015}. 

\citet{mcgovern2004} presented the first comprehensive infrared observations to reveal gravity-sensitive spectral signatures in young low-mass stars and brown dwarfs. The infrared and optical spectra of late-type giant stars and old field dwarfs were compared with the spectra of several young brown dwarfs to identify gravity-sensitive features, such as the \kii \ lines in the $J$-band, as well as TiO, VO, and FeH absorption systems. The paper also reported on the use of these spectral features to test the membership of potential very low mass brown dwarfs in young clusters. \citet{mcgovern2004} therefore forms the basis of the surface gravity analysis presented in this paper.

In this paper, we measure equivalent widths of \kii \ lines in the $J$-band and FeH absorption at 1.2 \microm \ for all targets in the BDSS, and use the A13 method to determine surface gravities for all objects for which the method is viable (spectral types M5-- L7). We expand upon previous surface gravity studies by calibrating the surface gravity classifications against objects of known ages from the literature, and discuss the extension of the gravity classifications beyond type L7. In Section 2, we discuss our observations and data reduction methods. Section 3 describes our method of determining surface gravity, and Section 4 discusses general trends,  interesting objects revealed by our analysis, and the ability of the gravity indices to distinguish the ages of objects. Section 5 summarizes our results.  

\section{Sample}

\begin{figure}
\begin{center}
\includegraphics[scale=0.46,angle=0]{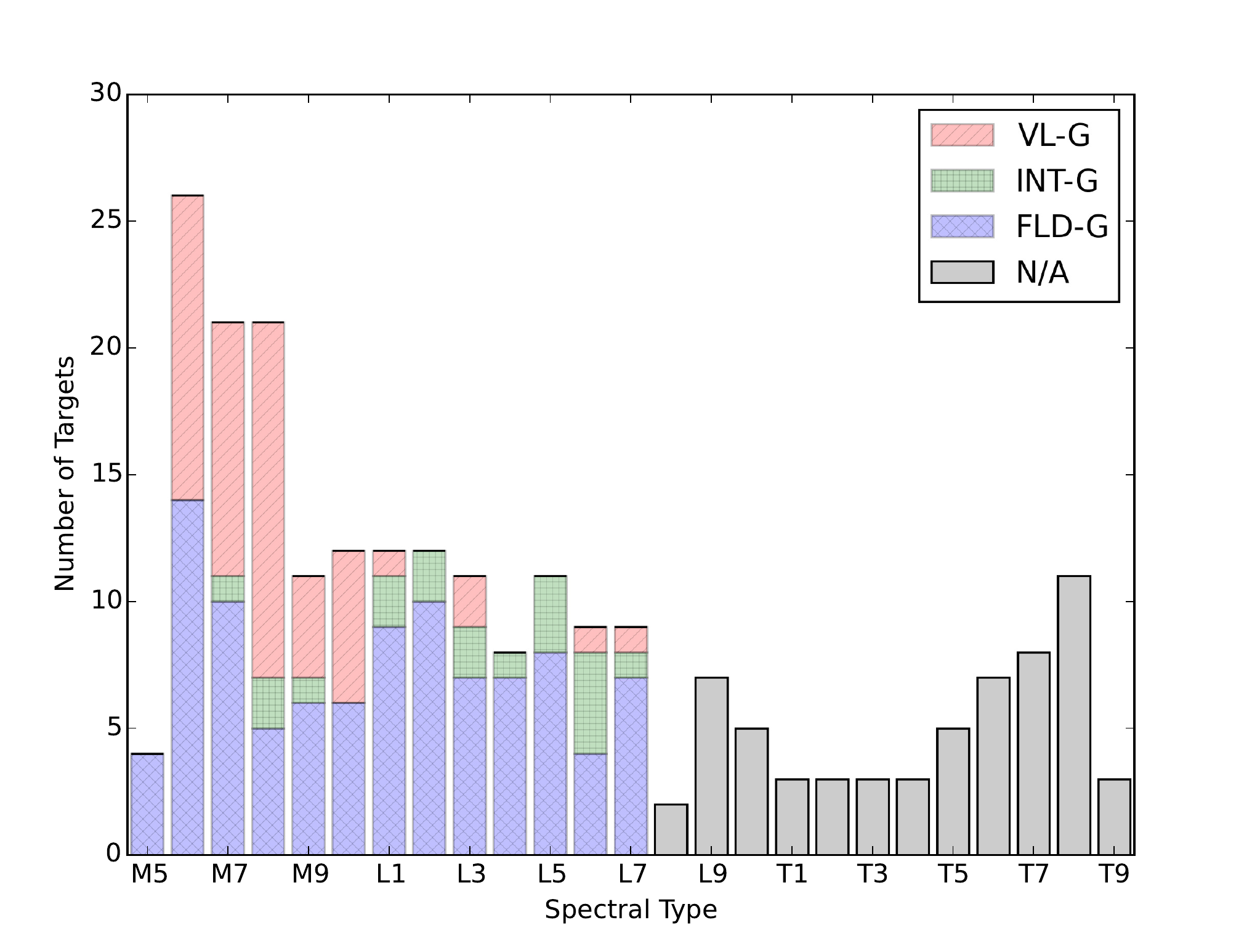}
\caption{Number of targets versus spectral type for all 228 objects in the BDSS. Shading represents gravity classifications, as defined by A13 and as determined in this paper. Red shading represents targets with a gravity classification of VL-G. These objects are likely very young. Green shaded regions denote targets with INT-G classification, indicating youth ($\sim$30--200 Myr). Blue targets have FLD-G gravity classifications, and are generally older than $\sim$200 Myr. Objects cooler than type L7 cannot be gravity typed by A13's methods and are shown in grey. 
\label{fig:num_vs_spt}}
\end{center}
\end{figure}

We present medium-resolution $J$-band spectra of 85 M dwarfs, 92 L dwarfs, and 51 T dwarfs, obtained as part of the BDSS. Ninety-seven spectra were published previously in \citet{mclean2003}, \citet{burgasser2003c}, \citet{mcgovern2004}, \citet{rice2010}, \citet{kirkpatrick2010}, \citet{luhman2012}, \citet{thompson2013}, Mace et al. (2013a)\nocite{mace2013a}, Mace et al. (2013b)\nocite{mace2013b} and \citet{kirkpatrick2014}, and the remaining 131 are presented here for the first time. By design, our sample spans a large range of spectral types, ages, and distances. In addition to known standards and field objects, we have observed peculiar objects such as  $J-K_{s}$ color outliers and known young and old objects. Sixty-four of our targets have age estimates based on their likely associations with clusters or moving groups, such as the Pleiades, Upper Scorpius, and Taurus regions, or from spectral analysis of stellar companions. However, the majority of the sample comprises field brown dwarfs and low-mass stars selected from the 2MASS and WISE surveys. As illustrated in Figure~\ref{fig:num_vs_spt}, our largest population of objects is late-type M dwarfs. 

\subsection{Observations}
Targets were observed using the strategy described in \citet{mclean2003} for the NIRSPEC instrument on the Keck II telescope in the non-echelle (medium-resolution) mode. For this mode, the slit used is typically 0.38$''$ wide (two pixels), though for several fainter T dwarfs and for observing conditions with sub-optimal seeing, the 0.57$''$ slit was used. For most observations, 300 s exposures were taken in nod pairs of 20$''$ separation along the 42$''$ slit. These nods were generally done in ABBA format for a total observing time of 20 minutes per target. Fainter objects were observed for longer, as needed. An A0V star at a similar airmass to each target was used for telluric corrections. If there were no nearby A0V stars, calibrators as early as B9  or as late as A3 were used instead. In the N3 filter ($\sim$1.15--1.35 \microm), the A0V stars typically only contain the Pa$\beta$ \ absorption line at 1.282 \microm \, which we interpolate over in the reduction process. In addition to telluric calibrators, flat field and dark frames were taken, as well as spectra of Ne and Ar lamps for wavelength calibration. Observation information for all targets in our sample is listed in Table~\ref{tab:obs}, as well as spectral types taken from the literature. \\

\subsection{Data Reduction}
All spectroscopic reductions were made using the REDSPEC package\textsuperscript{1}, software produced at UCLA by S. Kim, L. Prato, and I. McLean specifically for the reduction of NIRSPEC data as described in \citet{mclean2003}. The REDSPEC code first corrects for spatial and spectral distortion on the array using Ne and Ar lamp lines with wavelengths taken from the National Institute of Standards and Technology (NIST)\textsuperscript{2} \citep{nist}. Nod pairs of the target and calibrator are then background subtracted and divided by a flat field. Known bad pixels are removed as well. Spectra are obtained by summing over a range of $\sim$10 pixels (depending on seeing) and then dividing by the A0V calibrator spectrum to remove telluric features. Each pair of spectra was normalized and combined with other pairs (when available) to achieve a higher signal-to-noise ratio (SNR). This sample contains targets with SNR $\sim$10--200, though the majority of the spectra have SNR of at least 20. Finally, heliocentric velocity corrections were applied to the normalized spectra. We also performed a quality check on all data to ensure that wavelength dispersion solutions differed by less than $\sim 10^{-5}$ \AA /pixel. Plots and data files for all of our reduced spectra are available publicly through the BDSS archive\textsuperscript{3} or by request.

\footnotetext[1]{http://www2.keck.hawaii.edu/inst/nirspec/redspec.html}
\footnotetext[2]{http://physics.nist.gov/asd}
\footnotetext[3]{http://bdssarchive.org}

\section{Surface Gravity: Methods and Results}

Below we describe our method for calculating the EWs and spectral indices used to determine the surface gravities of our objects. We then present surface gravity estimates for all M6--L7 objects in the BDSS, as well as EW and spectral index values for all BDSS objects.

\subsection{Equivalent Widths}
\begin{figure}
\begin{center}
\includegraphics[scale=0.45,angle=0]{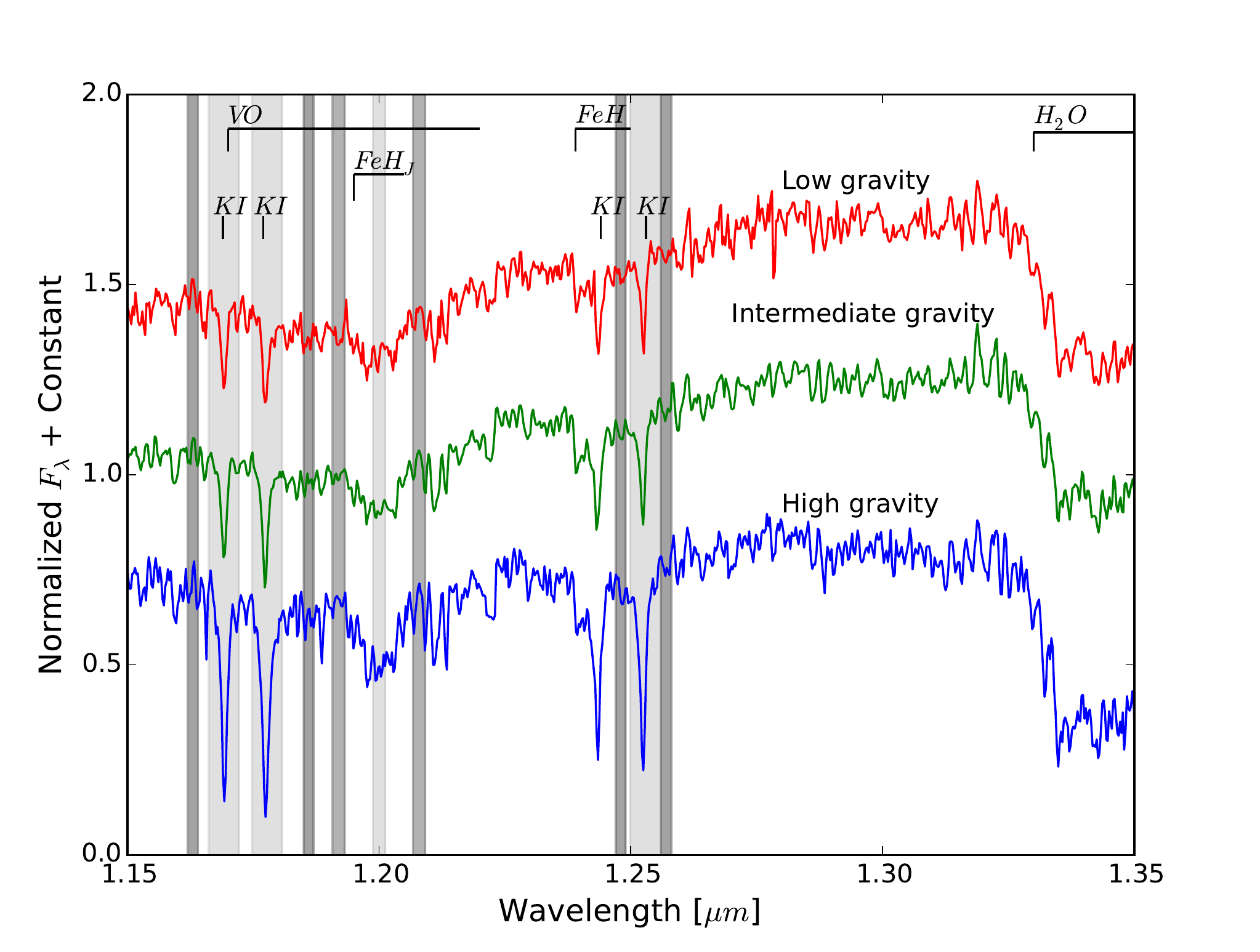}
\caption{Three example $J$-band spectra of spectral type L3 objects from the BDSS. Each spectrum represents a different gravity type: low gravity (2MASS 2208+2921), intermediate gravity (2MASS 1726+1538), and high gravity (2MASS 1300+1912). Major absorption features in the $J$-band are also labeled. Light grey shaded regions denote the locations used to calculate \kii \ EWs and the FeH$_J$ index. Dark grey regions denote locations of the pseudo-continua used in our calculations. The \kii \ line at 1.2437 \microm \ is marked, but not shaded. This line was not used to determine gravity types because of contamination from FeH absorption at $\sim$1.24 \microm .  
\label{fig:L3_example}}
\end{center}
\end{figure}

We compute pseudo-EWs for the four neutral potassium lines in the $J$-band following the method described in A13. For accurate comparison, we use the same line and pseudo-continuum windows as defined in A13 (see Figure~\ref{fig:L3_example}). The light grey shaded regions indicate the line windows used, while the dark grey shows the continuum windows. The \kii \ doublet at 1.1692 \microm \ and 1.1778 \microm \ share the continuum windows on either side of the doublet, and the 1.2437 \microm \ and 1.2529 \microm \ lines share the continua surrounding the 1.2529 \microm \ line. A13 chooses to exclude the 1.2437 \microm \ line from their final analysis because of the FeH contamination on the blue side of the line. For completeness, we compute and report EWs for this line. Indeed, we find that the 1.2437 \microm \ line exhibits more scatter and weaker correlation with surface gravity at this resolving power, and thus we also exclude it from our analysis.

Following a similar method to A13, we estimate a continuum value using a linear regression fit to the flux in the continuum windows. The EW calculations are performed using a Monte Carlo technique of 1000 iterations to estimate our uncertainties. Unlike A13, we do not use the rms scatter about the continuum fit to estimate the flux uncertainty. Instead, for each iteration of the Monte Carlo calculation, we modulate the flux in each pixel by adding a noise factor calculated by multiplying a random number drawn from a Gaussian centered at 0 with a sigma of 1, multiplied by the estimated noise determined by the SNR of that pixel. The equivalent width for each flux modulation is recorded, and we then compute the median and standard deviation of the EWs as the best estimate and 1$\sigma$ uncertainties. This is a similar method to the one described in \citet{aller2016}, who found that the method in A13 tended to underestimate flux errors in modest SNR spectra (SNR $\lesssim$ 200). 

We tested this technique using a range of number of iterations in our Monte Carlo calculations. We found that $\gtrsim$ 500 iterations were required to achieve stable results and that there is no significant difference between $10^3$ and $10^6$ iterations. In the interest of computational time, we opted for $10^3$ iterations.

\begin{figure*}
\begin{center}
\includegraphics[scale=0.475,angle=0]{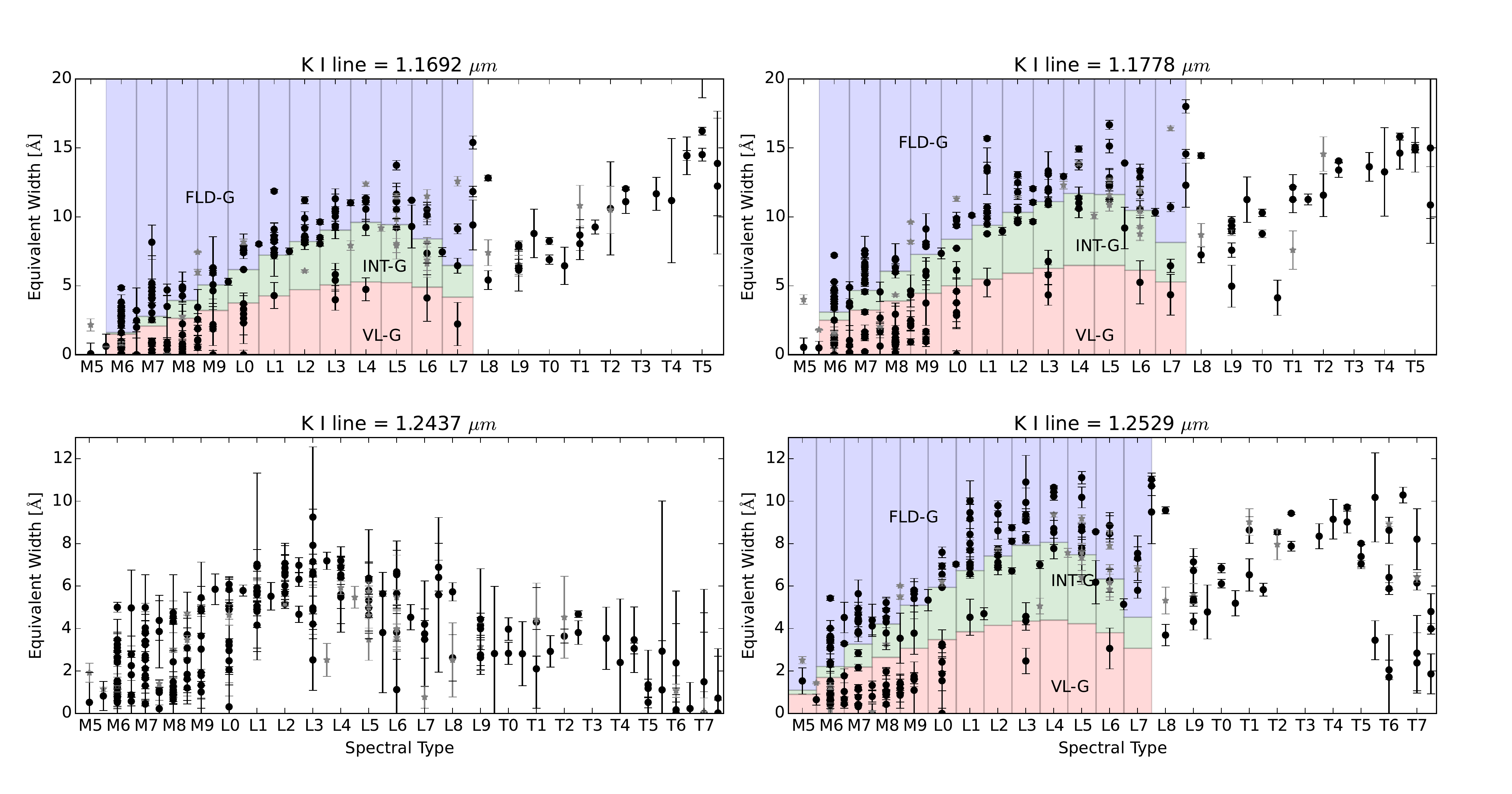}
\caption{\kii \ pseudo equivalent width vs spectral type for all M, L, and T dwarfs in the BDSS for which EWs can be measured. Field dwarfs are shown in black and binaries and subdwarfs are shown as grey stars. Uncertainties are calculated using a Monte Carlo technique with 1000 iterations of modulating the flux by the SNR and re-calculating the EW. The \kii \ lines at 1.1692 \microm \ and 1.1778 \microm \ disappear from T dwarf spectra later than $\sim$T5 and the \kii \ lines at 1.2437 \microm \ and 1.2529 \microm \ are not found in T dwarfs later than spectral type $\sim$T7. Shaded regions denote differing gravity types as defined by A13. Objects lying within the salmon shaded regions receive a score of ``2" (indicating low gravity), objects in the green shaded regions receive a score of ``1" (intermediate gravity), and objects within the blue shaded regions receive a score of ``0" (``field" or high gravity). These scores are used along with the FeH$_J$ score to compute a median gravity type. VL-G and INT-G designations are not distinguishable for M5 dwarfs for the \kii \ lines at 1.1692 and 1.1778 \microm , and gravity types are not designated for dwarfs of spectral type L8 and later. FeH contamination of the 1.2437 \microm \ line results in larger measurement uncertainties as well as a less-distinguishable low-gravity sequence. For this reason, A13 did not determine cutoff values for gravity types for this line.
\label{fig:EW_SPT_MLT}}
\end{center}
\end{figure*}

Table~\ref{tab:k1ew} lists our values for EW and uncertainties for the four \kii \ lines in the $J$-band for all objects in the BDSS. The first \kii \ doublet at 1.17 \microm \ disappears from the $J$-band spectra of dwarfs of spectral types $\sim$T5 and later. The \kii \ doublet at 1.25 \microm \ persists through $\sim$T7 (see spectral plots in Appendix). For this reason, objects later than T5 will have no \kii \ EW measurements at 1.17 \microm \ and objects later than T7 will have no \kii \ EW measurements at 1.25 \microm . 

In Figure~\ref{fig:EW_SPT_MLT}, we show results for the four \kii \ EWs versus spectral type for all M, L, and T dwarfs in the sample. Spectral types are taken from the literature (see Table~\ref{tab:obs}) and are measured in the NIR, if available. Shaded regions in Figure 3 show the boundaries proposed by A13 to designate low, intermediate, and high surface gravity objects, for objects of spectral type M6--L7.  

It should be noted that for some objects with apparently very low \kii \ absorption, the calculated EW can be less than zero. Visual inspection of these spectra shows that they do have very small or nonexistent \kii \ lines. In these cases, we have plotted these targets with EW values of zero, but the values listed in Table~\ref{tab:k1ew} are as measured. We believe this effect is because the EW calculation windows were chosen for objects with much deeper absorption lines. Objects with very weak \kii \ lines and a slightly higher continuum within the line-calculating region than the continuum region can thus have a negative EW. The negative EW does not affect the gravity classification of VL-G for these objects, so we chose to stay consistent with the A13 line and continuum boundaries when computing EWs. 

\subsection{FeH$_J$ Index}

In addition to the \kii \ equivalent width measurements, we studied the FeH$_J$ index from A13. This index measures the 1.2 \microm \ FeH absorption feature for medium-resolution (R$\sim$750--2000) data for objects of spectral type M6-L6. Figure~\ref{fig:L3_example} shows the window used for computing the index in light grey, and the windows used for estimating the continuum in dark grey. FeH absorption is found in late-type M dwarfs, most L dwarfs, and seen weakly in some T dwarfs. FeH absorption depth is known to correlate with surface gravity \citep{mcgovern2004}. Objects near the L/T transition do not show signs of FeH because the atmospheric conditions (i.e. cooler temperatures) have caused this molecule to precipitate (see, e.g. \citealt{marley2014}). Spectral types later than $\sim$ T1 show a slight re-emergence of the molecule \citep{burgasser2002b}, perhaps due to cloud-clearing, allowing flux to emerge from deeper layers within the brown dwarf, where some FeH remains in gaseous form (see also \citealt{tremblin2016} for an alternate interpretation). 

\begin{figure*}
\begin{center}
\includegraphics[scale=0.55, angle=0]{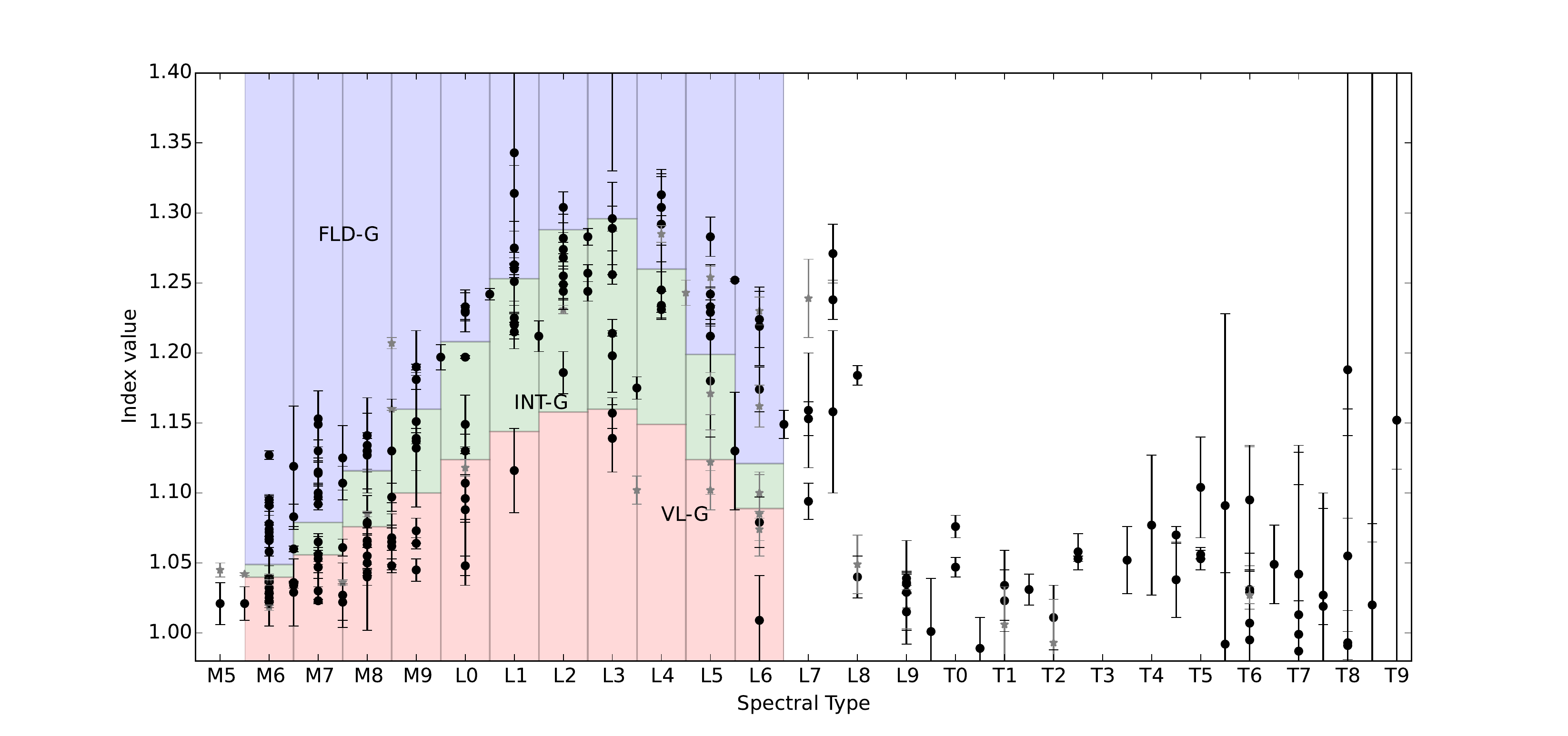}
\caption{FeH$_J$ index vs spectral type for all M, L , and T dwarfs in the BDSS. Normal dwarfs are shown in black and binaries and subdwarfs are shown as grey stars. Shading is the same as in Figure~\ref{fig:EW_SPT_MLT}. This index measures the FeH absorption feature at 1.2 \microm \ using the continuum and absorption bands shown in Figure~\ref{fig:L3_example}. FeH is found in late-type M dwarfs and most L dwarfs. Spectral types later than $\sim$ L8 have atmospheres cool enough to condense this molecule. Index values of $\sim$1.0 indicate the absorption feature is nearly absent in the spectra of L/T transition objects. This index re-emerges slightly in the mid-type T dwarfs. 
\label{fig:FeH_MLT}}
\end{center}
\end{figure*}

We present our FeH$_J$ index values for all objects in the BDSS in Figure~\ref{fig:FeH_MLT}. An index value of $\sim$1 is expected for the L/T transition dwarfs, indicating little to no absorption present in this spectral region. FeH$_J$ values for all BDSS targets are listed in Table~\ref{tab:k1ew}. 

\subsection{Gravity Scores}
A13 determined gravity score cutoff values for each \kii \ EW for spectral types M5--L7 and for the FeH$_J$ index for spectral types M6--L6 using a sample of known young and field objects. They assigned final gravity types using a median value of scores from multiple spectral indices across the 0.9 -- 2.5 \microm \ range, at both low and moderate resolutions. A13 gravity types for moderate resolution spectra are calculated from four scores determined by the following indices: VO (\textit{z}-band), \textit{H}-band continuum, FeH (either \textit{z} or \textit{J} band), and the mean score of the \nai \ and \kii \ EWs. As described above, we used similar methods and cutoff ranges to calculate our spectral indices and EWs. However, because we only had $J$-band spectra for most of our targets, we determined a gravity type using only the $J$-band medium-resolution indices from A13.

Gravity scores were computed for the \kii \ EWs at 1.1692 \microm , 1.1778 \microm , and 1.2529 \microm \ as well as for the FeH$_J$ index. To determine gravity scores for our sample, we compared our computed EW and FeH$_J$ values to the cutoffs tabulated in A13. If the index value was higher than the INT-G cutoff, it received a score of ``0", indicating field gravity. If the index value was between the INT-G and VL-G cutoffs, it received a score of ``1" and if the index value was smaller than the VL-G cutoff it received a value of ``2", indicating low surface gravity. Similar to \citet{aller2016}, we opted not to use the ``?" value, defined in A13, if the object receives a score that hints at intermediate gravity but has 1$\sigma$ uncertainties that overlap with field gravity values. These objects received a score of ``1".  We computed the median score from these four indices to determine the final gravity designation for each target. Following the method from A13, median scores less than or equal to 0.5 are classified as ``FLD-G", scores between 0.5 and 1.5 are classified as ``INT-G", and scores greater than or equal to 1.5 receive ``VL-G" classification. Table~\ref{tab:k1ew} lists gravity scores and the resultant gravity classification for objects of spectral type M5-L7. When available, the gravity score given by A13 is also listed.

Using multiple indices to characterize the surface gravity allows some objects to be seen as having borderline gravity classifications between VL-G and INT-G or INT-G and FLD-G. The combination of multiple indices is more robust against any particular index skewing the classification. Errors in an index might come from measurement errors or from physical effects causing the absorption in one index to be abnormal compared to the other indices calculated for a particular target. Sixty-two objects out of the 159 for which A13 gravity types were computed had more than one type of score. However, only 7 targets received individual index scores spanning all three gravity types, and three of these objects are binaries or subdwarfs (see \S~\ref{sec:excluded}).

\subsection{Radial Velocity}

One consideration we made in our calculations was the effect of radial velocity (RV) on the EW measurements and therefore on the gravity estimations. One resolution element in the medium-resolution mode of NIRSPEC is equivalent to 150 \kms. Though rarely occurring, high RV targets could have their spectra shifted by a large enough amount that the calculation of a gravity estimate would be significantly altered. In order to understand the effect of RV on our EW and spectral index values, we examined 21 objects in our sample with known RVs from the literature, with RV magnitudes ranging from $\sim$5 \kms \ to 195 \kms. First, we shifted their spectra to account for the known RV offset. Then, we recalculated their \kii \ EWs and FeH$_J$ indices and gravity types, and compared these values to our original calculations. Only two of our targets, the known L subdwarfs 2MASS 0532+8246, SDSS 1256-0224 had RVs in excess of 100 \kms (\citealt{burgasser2003c, burgasser2009}, respectively). The other 19 targets had RVs $\lesssim$ 30 \kms. 

None of our RV-shifted targets had differing gravity types from our original calculations. \kii \ EW and FeH$_J$ index values differed by less than $\sim$5 \% for all of the targets. Few dwarfs have measured RVs in excess of 200 \kms \ as the majority belong to the disk population and have similar space motions to the Sun. Because non-echelle NIRSPEC spectra can only resolve radial velocities greater than 150 \kms \ without cross-correlating to known RV targets, we estimate that this has a minimal impact on our measured gravity types. 

We performed additional analysis to test the effect of RV on the EW measurements by performing a Monte Carlo simulation of 1000 iterations on high SNR spectra of both field age and young targets, each time drawing a random RV from a normal distribution with $\sigma_{RV}$ of 100 \kms \ and re-calculating the EW. The resulting median and standard deviation of the distribution were entirely consistent with our original measurements. We therefore conclude that the RV of the target does not influence these calculations.

\subsection{Excluded Objects} \label{sec:excluded}
We present J-band spectra and measure equivalent widths and FeH absorption for all BDSS targets, where relevant. However, two sub-populations of our sample were removed from the surface gravity analysis: known binaries and subdwarfs, whose spectral features are known to vary from the general field population for reasons other than their surface gravity. 

\subsubsection{Binaries} \label{sec:binaries}

LP 213-67 (M8+L0; \citealt{close2003}),  2MASS 0850+1057 (L6+L7; \citealt{reid2001, burgasser2011a}), SDSS 0805+4812 (L4+T5; \citealt{burgasser2007, burgasser2016b}), 2MASS 2140+1625 (M8.5+L2; \citealt{close2003}), 2MASS 2152+0937 (L6+L6; \citealt{reid2006}), 2MASS 1315-2649 (L3.5+T7; \citealt{burgasser2011c}) are known spectral binaries in our sample. We caution against inferring a gravity type or age estimate for these objects, as their combined spectra could have an effect on the gravity-sensitive indices. For example, 2MASS 1315-2649 (L3.5+T7), which \citet{burgasser2011c} finds to be at least 1 Gyr old given its kinematics, has an INT-G gravity type, which would imply an age of $\lesssim$ 100 Myr. It is possible that this discrepancy in age estimate is caused by binarity.

\subsubsection{Subdwarfs}
Four targets in our sample are known subdwarfs, LHS 1135 (d/sd M5; \citealt{kirkpatrick2010}), WISE 0435+2115 (sd L0; \citealt{kirkpatrick2014}), SDSS 1256-0224 (sd L3.5; \citealt{burgasser2009}), and 2MASS 0532+8246 (sd L7; \citealt{burgasser2007}). These objects tend to have large space motions, are typically found to be part of the Galactic halo population, and generally have sub-solar metallicity, although they exhibit stronger hydride features than similarly classified dwarfs. Because of their low metal content, we chose to exclude these objects from our analysis and do not determine gravity types for the subdwarfs in our sample. It should be noted that subdwarfs can exhibit small \kii \ EWs due to their lower metal content. These smaller EWs can be misleading as it is thought that these objects are quite old, and should not exhibit signs of low gravity. For example, the red \kii \ doublet in the J band of SDSS 1256-0224 is weak enough to infer low gravity, though the strength of its FeH$_J$ index implies high gravity and as a subdwarf it is likely older than $\sim$ 5 Gyr. \\ \\


\section{Discussion}

\subsection{Comparison to \citet{allers2013}}

\begin{figure*}
\begin{center}
\includegraphics[scale=0.6,angle=0]{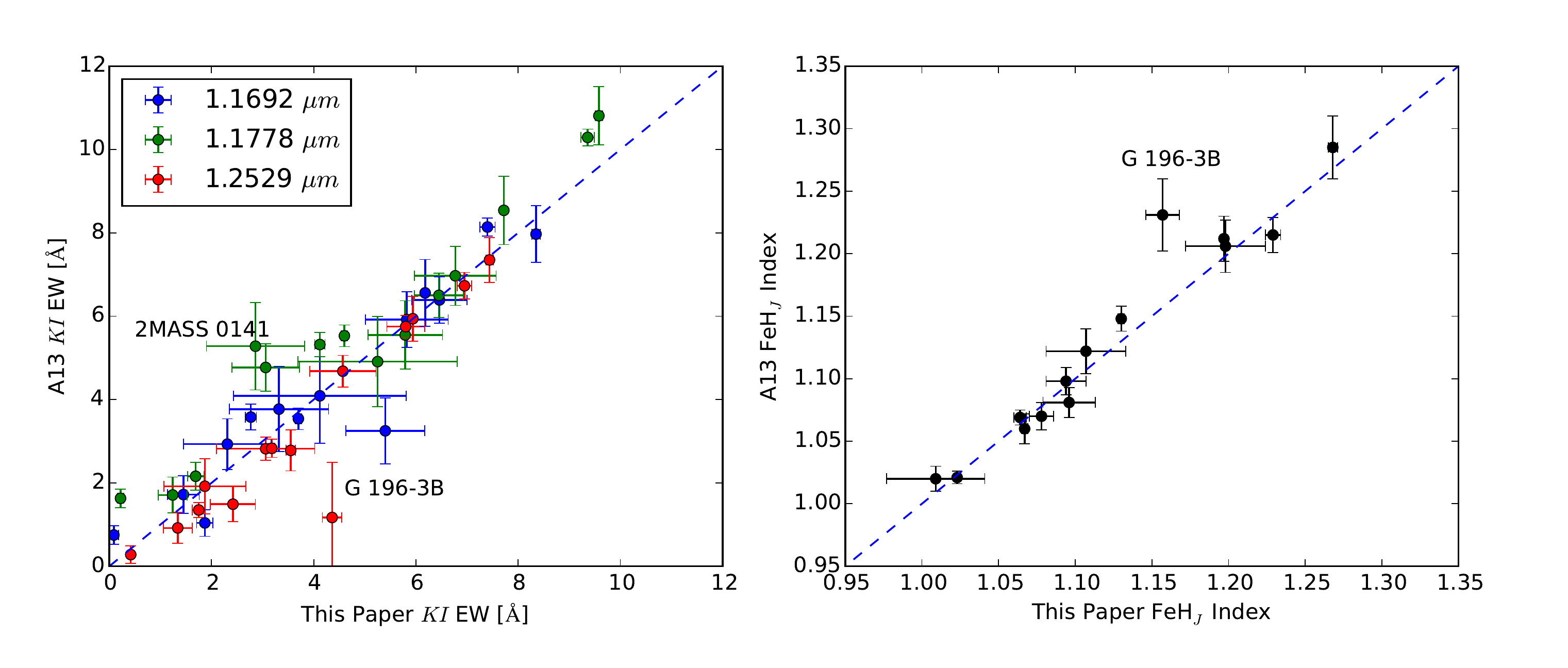}
\caption{Left: comparison of \kii \ EW values between A13 and this paper. Colors indicate the particular line at which the EWs were calculated. The one-to-one line is shown to aid comparison. Right: comparison of FeH$_J$ index values from A13 and this paper. Despite differing instruments and resolving powers, our values are consistent within the uncertainties. 
\label{fig:allers_em}}
\end{center}
\end{figure*}

In Figure~\ref{fig:allers_em}, we plot our \kii \ EW values (left) and FeH$_J$ index values (right) versus those of A13, for the overlapping targets in our samples. We find that although the two data sets use different instruments with different resolving powers, our results are consistent within the uncertainties. The 1.1778 \microm \ EW values appear to be slightly higher on average in A13 than in our own analysis, but our values are consistent within 2$\sigma$. The 1.2529 \microm \ line appears to have the opposite result, with our values being slightly higher than those presented in A13. The major outlier is G196-3B, which has a lower SNR spectrum in A13, as indicated by its larger error bar. A13's value is less than 2$\sigma$ away from our result.

We find that the modified technique using only \textit{J}-band indices with NIRSPEC R $\sim$ 2000 spectra produces consistent results to the gravity classifications determined using spectral indices across the \textit{z, J, H,} and \textit{K} bands. Of the 20 matching targets between the two samples, all targets except one receive the same designation as found by A13, allowing for overlap in the borderline designations. For example, A13 finds that PC0025+0447 has intermediate gravity, while we classify it as borderline VL-G/INT-G. The exception is GL 417 BC (L4.5+L6) which we exclude because of its binary nature. When compared to the index value cutoffs for an L5 dwarf, we designate this object as FLD-G, as does A13.

\subsection{Overall Trends}
Before discussing overall trends in our sample, we must clarify that all spectral types for our objects were compiled from the references listed in Table~\ref{tab:obs}. Some objects were classified in the red-optical, while others were classified in the NIR, and objects can have a spectral type uncertainty as large as $\pm$2 spectral types. Such discrepancies have been well documented in the literature and several methods have been presented in various papers for determining spectral types. For this reason, we can expect that the uncertainty in spectral type will cause larger uncertainty in the the overall trends.

As noted in \citet{burgasser2002b} and \citet{mclean2003}, potassium equivalent widths in the $J$-band tend to rise with increasing spectral type from M5 to $\sim$L5 and at $\sim$L5--L7 the EWs drop, rising again with increasing spectral type around L8. We see this same trend in the full sample (Figure~\ref{fig:EW_SPT_MLT}), though it shows a large amount of scatter. Objects lying below the trend exhibited by the field dwarfs are primarily those exhibiting signs of youth (VL-G and INT-G). However, some objects show signs of low gravity in one absorption feature, while exhibiting field-like features elsewhere in their spectrum. For this reason, gravity types should be based on multiple gravity-sensitive indices, as A13 also cautions.

The behavior of the FeH$_J$ index follows a similar trend to the \kii \ EWs (Figure~\ref{fig:FeH_MLT}), although the FeH$_J$ index peaks at $\sim$L3, drops out almost entirely near the L/T transition, and then re-emerges at much lower levels of absorption in the mid-T spectral types, before again dropping out almost entirely in the late T's. This trend is similar to results seen by \citet{burgasser2003b} for the FeH feature at 0.9896 \microm . They note a weakening in FeH band strength in late-type L dwarfs followed by a slight strengthening, near spectral type $\sim$T5.5, before disappearing again. \citet{burgasser2003b} proposes that the re-emergence of this feature in the T dwarfs is an indication of cloud clearing. Holes in the cloud deck or a complete lack of clouds in the upper atmosphere allow the observer to detect light from deeper within the atmosphere of the brown dwarf, where the temperatures are warm enough to sustain the presence of the FeH molecule. This interpretation has recently been challenged by \citet{tremblin2016}, who find that the FeH reversal can be reproduced by thermochemical instability effects, rather than cloud opacity changes. Regardless of the interpretation, we verify the trend in the re-emergence of FeH absorption.

The A13 gravity classifications do not extend to spectral types cooler than L7. We are unable to extend these classifications to later spectral types, even with our larger sample. Establishing a low gravity sequence requires a large enough sample of field dwarfs to determine the field sequence. Additionally, a large sample of known young objects are required to determine the location of the low gravity objects. Currently there are very few known young late-type L or early T dwarfs, none of which are in our sample. Searches for very low mass objects in nearby young moving groups could yield a larger sample to carry out such a study, but this is not possible with the sample presented here. 

Some of the VL-G and INT-G objects in our sample are known ``red" L dwarfs because their $J-K_{s}$ colors are significantly redder than the $J-K_{s}$ colors of typical field dwarfs. Red $J-K_{s}$ color can be an indication of youth, though the term ``red" should be reserved for those L dwarfs with red $J-K_{s}$ color that do not otherwise show signs of youth (see \citealt{kirkpatrick2010} for further discussion of the red and blue L dwarfs). Likewise, L dwarfs with significantly bluer $J-K_{s}$ colors compared to typical field dwarfs are called ``blue", though this nomenclature should also be reserved for those L dwarfs with significantly bluer colors that do not exhibit signs of very low metallicity. In general, we find that the ``red" and ``blue" L dwarfs do not show consistent signs of low or high gravity, respectively.

\subsection{Comparison of Objects of Known Age}
\begin{figure*}
\begin{center}
\includegraphics[scale=0.45,angle=0]{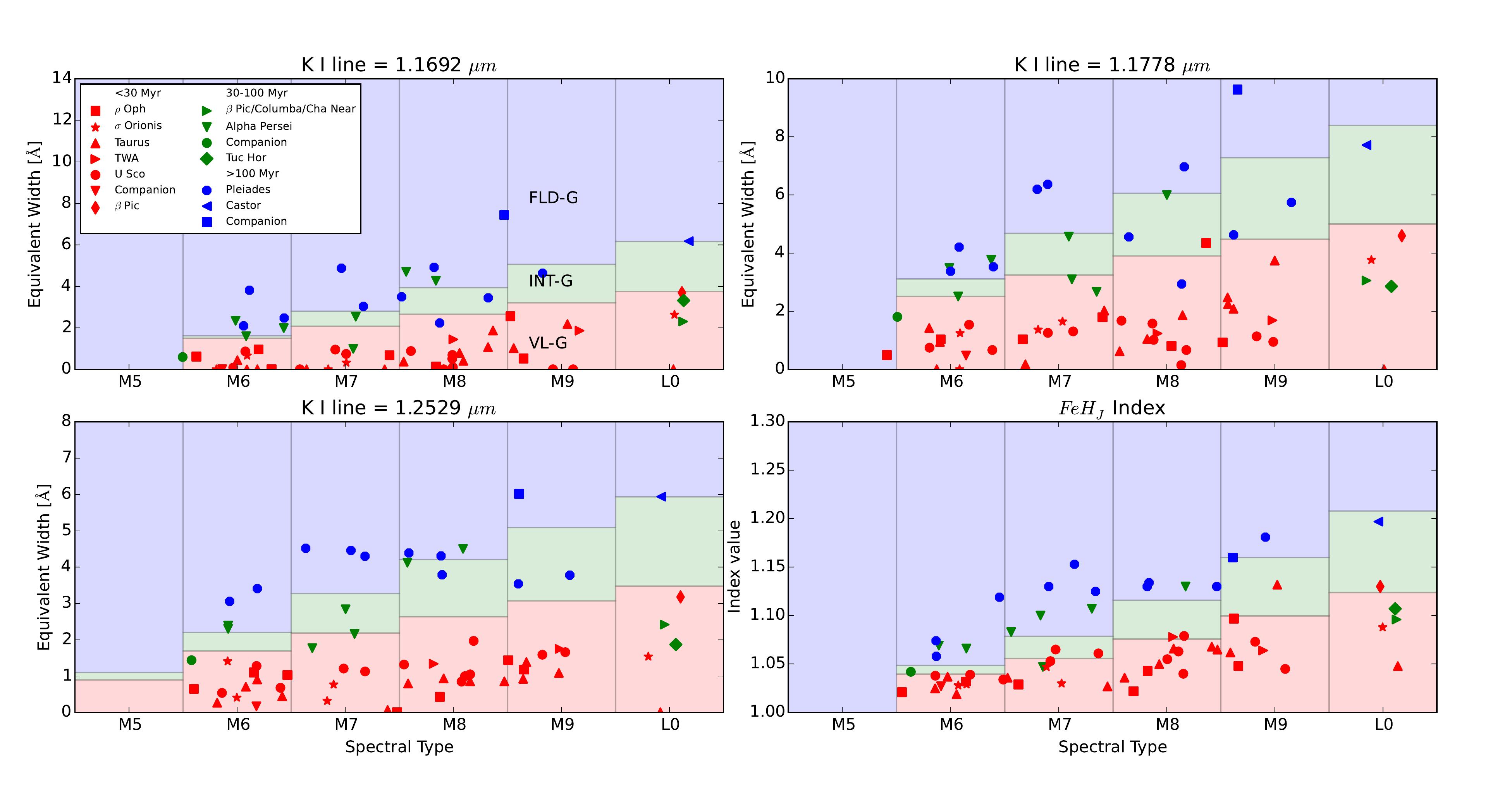}
\caption{EW vs SpT and FeH$_J$ vs SpT for dwarfs of spectral type M5.5-L0 with known ages. Different shaped symbols represent the methods used to estimate ages, i.e. group membership or an age estimate from a more massive stellar companion. For references, see Table~\ref{tab:ages}. Symbols are colored by their known ages as follows: Red symbols have ages $<$ 30 Myr, green symbols are $\sim$30 -- 100 Myr, and blue symbols are $>$100 Myr. Binaries from Table~\ref{tab:ages} are not shown, nor is the only single object of spectral type later than L0 in our sample, 2MASS 2244+2043 (L6, VL-G, AB Dor). Spectral types have been distributed randomly in each spectral type bin for ease of viewing.
\label{fig:ages_M}}
\end{center}
\end{figure*}

To understand the age limits represented by the gravity classifications, we compare objects with known or predicted ages in the literature, determined by independent methods, such as kinematics or companionship to a well characterized star. Table~\ref{tab:ages} lists age estimates and gravity types for 64 objects in the BDSS with previously determined ages. All BDSS targets that are candidate or suspected members of nearby young associations are included. Likelihood of membership (where available in the literature) is noted as well. Also included are several targets with age estimates from their more massive companions. Figure~\ref{fig:ages_M} shows the three adopted gravity-sensitive \kii \ EWs and the FeH$_J$ index versus spectral type for objects with known ages for dwarfs of spectral type M5.5-L0.  Binaries from Table~\ref{tab:ages} are not shown, nor is the only single object of spectral type later than L0 in our sample of known-age objects, 2MASS 2244+2043 (L6, VL-G, AB Dor). Red symbols represent objects of ages $<$ 30 Myr, green symbols denote objects between $\sim$30 and 100 Myr, and blue symbols represent objects $>$100 Myr. Varying shapes are used to distinguish between the different young associations (see figure legend for more details). 

As seen in Figure~\ref{fig:ages_M} and Table~\ref{tab:ages}, members of various associations tend to have the gravity type corresponding to the estimated age of the association. A few targets have previously been found to be interlopers, so we exclude these from our analysis. Additionally, several objects with known ages are also tight binaries with potentially contaminated spectra (see \S~\ref{sec:binaries}). Binaries of known ages are listed in Table~\ref{tab:ages}, but are excluded from Figure~\ref{fig:ages_M} and any additional age calibration analysis (\S~\ref{sec:ages}). Below we discuss each of the associations in order of estimated age.

All six of the $\rho$ Ophiuchi candidates ($<$1 Myr, \citealt{greene1995}), all thirteen Taurus ($\sim$ 1.5 Myr, \citealt{briceno2002}) candidates, and 2MASS 2234+4041 (1 Myr; Companion to LkH$\alpha$ \ 233; \citealt{allers2009}), have VL-G classifications, as expected. For targets with $\sigma$ Orionis (3 Myr, \citealt{brown1994}) designations, all but one of the six targets is classified as VL-G. Only S Ori 47 is classified as INT-G (not shown in Figure~\ref{fig:ages_M} or Table~\ref{tab:ages}). Unlike the other $\sigma$ Orionis candidates, this target's $J$-band spectrum shows very clear \kii \ absorption features, more akin to a field dwarf, as noted in \citet{mcgovern2004}. \citet{mcgovern2004} conclude that this object is likely a several Gyr old object located $\sim$120 pc away, with a mass near the hydrogen burning limit, and is not associated with the $\sigma$ Orionis cluster. Our analysis of S Ori 47 suggests that S Ori 47 is likely much younger than 1 Gyr, but certainly older than $\sim$30 Myr and very unlikely to be associated with $\sigma$ Orionis. Based on the age vs. EW values in Section~\ref{sec:ages}, we estimate that S Ori 47 has an age closer to $\sim$150 Myr, and is likely an intermediate aged field dwarf.  For this reason, we have excluded S Ori 47 from the age-calibrated sample in \S~\ref{sec:ages}.

 Both TW Hya ($\sim$10 Myr, \citealt{bmn2016}) targets have VL-G classifications. Of the fifteen objects with Upper Scorpius (11 $\pm$ 2 Myr, \citealt{pecaut2012}) designations in our sample, all but three have VL-G classifications. U Sco 121, 85, and 132 each have FLD-G designations, and they are all previously suspected non-members \citep{muzerolle2003}. Our analysis supports this conclusion. These three objects are not shown in Figure~\ref{fig:ages_M} or Table~\ref{tab:ages} and are excluded from the age-calibrated sample in \S~\ref{sec:ages}. The $\beta$ Pic (21--26 Myr; \citealt{bmn2016}) target, 2MASS 0443+0002, is classified as VL-G. 

2MASS 0141-4633 in Tucana Horlogium (45 $\pm$4 Myr, \citealt{bmn2016}) is classified as VL-G. 2MASS 0608-2753 also receives a VL-G classification. Based on results from \cite{gagne2014} and \cite{faherty2016}, we list this target as a candidate member of three groups: Cha-Near ($\sim$10 Myr, \citealt{zuckermansong}), $\beta$ Pic, and Columba ($42\substack{+6 \\ -4}$ Myr, \citealt{bmn2016}).

Of the Alpha Persei (80--100 Myr, \citealt{stauffer1999}) members, one is classified VL-G, one is INT-G, and the other five are FLD-G. AP 270, which receives a VL-G classification, is less likely to be a member of Alpha Persei and could potentially be a young interloper. Gl 577 BC (70 Myr; companion) is classified as FLD-G. 

The majority of the Pleiades ($\sim$125 Myr) targets receive FLD-G classifications. Two Pleiades members (Roque 7 and Roque 4) are classified as INT-G, and Teide 1 is a borderline VL-G/INT-G object, but all three of these targets have much lower SNR spectra (SNR $\sim$10) and thus have much more uncertain gravity types. \citet{simon2006} used a comparison to the Pleiades to date Gl 569 BC at $\sim$100 Myr. Gl 569 BC receives a FLD-G classification using our method, similar to the Pleiades objects studied here.

Surprisingly, the AB Doradus ($149\substack{+51 \\ -19}$ Myr, \citealt{bmn2016}) candidate 2MASS 2244+2043 is classified as VL-G in our analysis. Several other studies of the members of AB Doradus have determined a variety of gravity classifications for different members. A13 and \citet{faherty2016} also present AB Doradus members with VL-G signatures, as well as members with INT-G and FLD-G classifications. \citet{aller2016} presents new AB Doradus members with INT-G classifications.

 LP 944-20 has been identified as a member of the Castor moving group (400 $\pm$40 Myr, \citealt{zuckerman2013}), though the existence of the group is disputed and age estimates vary broadly for proposed members. (See e.g., \citealt{monnier2012} and \citealt{mamajek2013}). However, LP 944-20 also has a Li measurement in \citet{reiners2009}, suggesting an age $<$ 500 Myr and implying that this target is younger than the typical ``old" field dwarf. This target receives a FLD-G designation. 

The object in our sample with the oldest measured age is Gl 417BC, which \citet{allers2010} estimate to be $750\substack{+140 \\ -120}$, based on gyrochronology. \citet{kirkpatrick2001a} estimates an age of 80--300 Myr based on various dating methods for Gl 417A. This target is also a FLD-G object.

Our analysis suggests that the VL-G classification is only sensitive to ages as old as $\sim$20--30 Myr, as originally proposed in A13. The INT-G designation appears to probe only the $\sim$30--100 Myr range, while the FLD-G designation probes $\gtrsim$100 Myr, not $\gtrsim$200 Myr as suggested by A13. However we see that, similar to the results seen in \citet{faherty2016}, there is a spread in gravity classifications even amongst targets belonging to the same association although they are assumed to be coeval. In Section~\ref{sec:ages} we further examine the age ranges probed by each gravity designation as a function of spectral type and \kii \ line. \\

\subsection{Potentially Young Objects}

Here we highlight targets with VL-G and INT-G designations that are not previously discussed in A13, and are not known members of nearby young associations or young clusters. For each of the targets, we calculate the BANYAN II v1.4 likelihood of membership in various nearby young moving groups, as well as likelihood of being a young ($<$1 Gyr) or old ($>$ 1 Gyr) field object \citep{gagne2014, malo2013}. The BANYAN II tool utilizes the 3D space motions and positions of many nearby young moving groups to determine via bayesian statistics the likelihood of a target being a member of a nearby young association. Not all associations are accounted for, so a BANYAN II ``young field" object could be a member of a young association not included in BANYAN II, or it could indeed be a young field dwarf, that is, a field dwarf exhibiting signs of youth. BANYAN II requires at least target coordinates and proper motion to estimate membership probability, but we input distance and radial velocity information for the BANYAN II online tool when available from the literature. We used the priors developed by and outlined in \cite{gagne2014} and did not use the uniform priors option.

\textit{2MASS 1459+0004} is an M6 dwarf with a VL-G designation. \citet{kirkpatrick2010} present the discovery of this object as well as a proper motion of \mua = 308 $\pm$ 248 \masyr and \mud = -342 $\pm$ 275 \masyr. BANYAN II results for this target suggest $\lesssim$ 1 \% likelihood of this object belonging to Argus or AB Dor, a 13.9 \% probability of being a young field object, and 85.85 \% likelihood of being an old field object, based solely on the target's coordinates and proper motion. If we assume the target is $<$ 1 Gyr old, it then receives a 98.2 \% probability of being a young field object.

\textit{2MASS 1331+3407} in an L1pec object with an INT-G classification noted as being particularly red by \citet{kirkpatrick2010}. \citet{gagne2014} found that this object has no likelihood of belonging to a nearby young moving group, so this is most likely an intermediate-aged field dwarf. Having particularly red spectroscopic or photometric features can be an indication of youth, though \citet{kirkpatrick2010} emphasizes that the term ``red" should be reserved for objects with significantly red $J-K_{s}$ colors or spectra that do not show signs of youth.

\textit{2MASS 0543+6422} is an L2 dwarf with an INT-G classification. \citet{gagne2015b} also gave this object an INT-G classification based on an IRTF SpeX spectrum and do not find any probability of this object belonging to currently known young moving groups.

\textit{2MASS 1841+3117} is an L4pec dwarf with an INT-G designation. The optical spectrum for this object in \citet{kirkpatrick2000} is noticeably blue, which can imply higher gravity, however the \kii \ lines in the J band exhibit signs of lower gravity. It is possible that the peculiar nature of its spectrum is implying that a physical mechanism other than low-gravity could be the cause of the smaller \kii \ EWs, or that its blueness could be caused by some reason other than high surface gravity. BANYAN II results using coordinates, proper motion, and parallax for this object \citep{faherty2009} suggest a 54 \% probability that this object is a young field object.

\textit{2MASS 1553+2109} is an L5.5 dwarf with an INT-G classification. This object is known to have red NIR colors and strong Li absorption \citep{kirkpatrick1999}, a further indication that it is a young field dwarf. Based on the BANYAN II results using the kinematics from \citet{schmidt2010}, this object has 30.5 \% likelihood of being a young field dwarf, and a 69.5 \% likelihood of being an old field dwarf.

\textit{2MASS 0740+2009} is an INT-G classified L6 dwarf, previously found to have unusually red $J-K_{s}$ colors \citep{thompson2013}. Its red colors could be attributed to lower surface gravity, in this case, and is likely younger than $\sim$ 100 Myr. Using the kinematics and distance from \citet{faherty2009} and the the BANYAN II predictions, we find only a 2.4 \% likelihood that this object belongs to the young field population and a 97.6 \% likelihood of being an old field dwarf. 

\textit{2MASS 2151+3402} is an L7pec dwarf with a VL-G classification. However, \citep{kirkpatrick2010} find that it has slightly blue NIR colors. This particular object has a low SNR spectrum and it is likely that the noise contaminated the estimation of the \kii \ EWs. We smoothed the spectrum using a Gaussian 1D Kernel with a width of 3 pixels and re-calculated the \kii \ EWs using the methods described above. After smoothing, the gravity scores this object receives are ``1", ``1", and ``0".  Additionally, if FeH$_J$ were defined for L7 dwarfs, this would likely receive a FLD-G designation for that index, making it more likely a FLD-G object overall. \citet{schneider2014} publish a measurement of the H$_2(K)$ index for this object, which they measure to be larger than the median H$_2(K)$ value for L7 dwarfs. The H$_2(K)$ index \citep{canty2013} is an index designed to measure the slope of the K-band continuum, which is known to be ``peakier" in low-gravity objects. A high H$_2(K)$ for this object is further indication that this is likely to be a field-gravity object. However, BANYAN II predictions based on the proper motion from \citet{kirkpatrick2010} and the sky coordinates suggest a slight ($<$ 1\%) probability that this object could be a member of $\beta$ Pic or Columba, a $\sim$1--2 \% probability of belonging to Argus or AB Dor, and a 34.4 \% likelihood of being a young field object.

\subsection{Determining Ages \label{sec:ages}}

\begin{sidewaysfigure*}
	\centering
	\vspace{3.5in}
	\includegraphics[scale=0.8,angle=0]{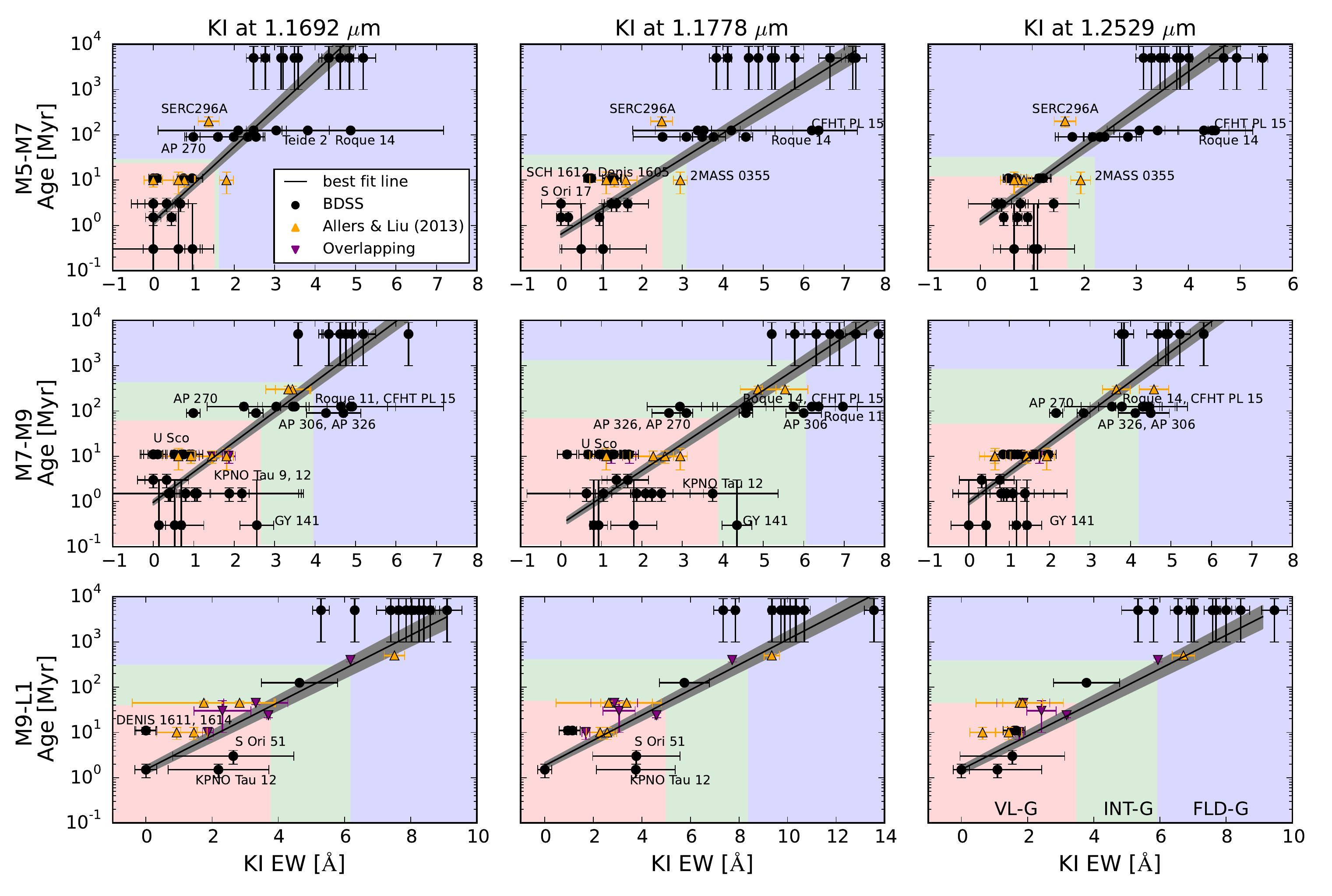}
	\caption{Age vs \kii \ EW at 1.1692 \microm, 1.1778 \microm, and 1.2529 \microm, binned by spectral type. Blue, green, and red shaded regions are the same as Figure~\ref{fig:EW_SPT_MLT}. BDSS objects with known ages are shown in black circles, alongside a field sample selected from targets with ``FLD-G" designations and given ages of 5 $\pm$ 4 Gyr to bound the upper limit of age as a function of EW. To increase our sample size we include objects from A13 with moderate resolution data and known ages (yellow upward-facing triangles). Objects with overlapping data between the A13 sample and our sample are marked by purple downward-facing triangles. Significant outliers are denoted in each panel as well. The black line represents the best-fit line as determined by a weighted orthogonal distance regression using the \textit{scipy.odr} package in python. The grey shaded region represents the 1$\sigma$ uncertainties in both slope and intercept.
	\label{fig:age_vs_ew}}
\end{sidewaysfigure*}

To further investigate the ability of the gravity indices to determine ages, we study the dependence of object age, taken from the literature, with \kii \ EW for the lines at 1.1692 \microm , 1.1778 \microm ,  and 1.2529 \microm \  binned by spectral type (Figure~\ref{fig:age_vs_ew}). There are a total of 73 objects used in this analysis, which are not known to be binaries and are assumed to be reliable age calibrators. Of these, 51 objects are BDSS targets, 15 objects come from A13, and 7 objects overlap both the BDSS and A13 samples. An additional 24 BDSS targets without known associations to young moving groups that received FLD-G designations in all four indices are also plotted, with age estimates of 5$\pm$4 Gyr. Each panel plots age vs. \kii \ EW for a bin of three spectral types, because of the need to remove the previously shown trend of EW with spectral type. In general, spectral types are only known to $\pm$ 1 type, so this coarser grouping of spectral types is analogous to the inherent spread in spectral features seen by objects of the same given spectral type. There is a clear linear trend between the \kii \ EWs and log(Age) as displayed in Figure~\ref{fig:age_vs_ew}. Thus, for each graph we perform a weighted orthogonal distance regression to determine a best fit function of the form in Equation~\ref{eq:age_vs_ew} using the \textit{scipy.odr} package \footnote{https://docs.scipy.org/doc/scipy/reference/odr.html} in python.

\begin{equation} \label{eq:age_vs_ew}
Age = A \times 10^{B  \times EW }
\end{equation}

Parameters and 3$\sigma$ uncertainties for the best fit lines for each spectral type bin from M6 to L0 and each \kii \ line are listed in Table~\ref{tab:ew_age}. For ages in units of Myr and EWs measured in \AA, the coefficient A is in units of Myr and B is \AA$^{-1}$. 

Figure~\ref{fig:age_vs_ew} shows only three of the spectral type bins. The figures for spectral types M7 $\pm$ 1 and M9 $\pm$ 1  show similar trends and are not shown, however the best fit parameters for these spectral type bins are listed in Table~\ref{tab:ew_age}.We were unable to achieve satisfactory fits to the function of log(Age) vs. FeH$_J$ index, likely because the range of values for the FeH$_J$ index is much smaller. Additionally, we were unable to extend the age vs. EW relationship beyond $\sim$L0 because of the lack of later-type, age-calibrated objects in our sample. 

The red, green, and blue shaded regions in Figure~\ref{fig:age_vs_ew} are taken from the A13 boundaries for VL-G, INT-G, and FLD-G for the average spectral type at each wavelength. These aid in demonstrating the large and varying age ranges probed by each of the gravity types. From this figure, one can see that the large inherent spread in EW value makes it difficult to draw firm conclusions about the age of an object solely based on the measurement of its \kii \ EWs. Coeval objects of similar spectral types can have widely varying EWs, as mentioned previously in regards to the AB Doradus moving group. 

\begin{figure}
\begin{center}
\includegraphics[scale=0.45,angle=0]{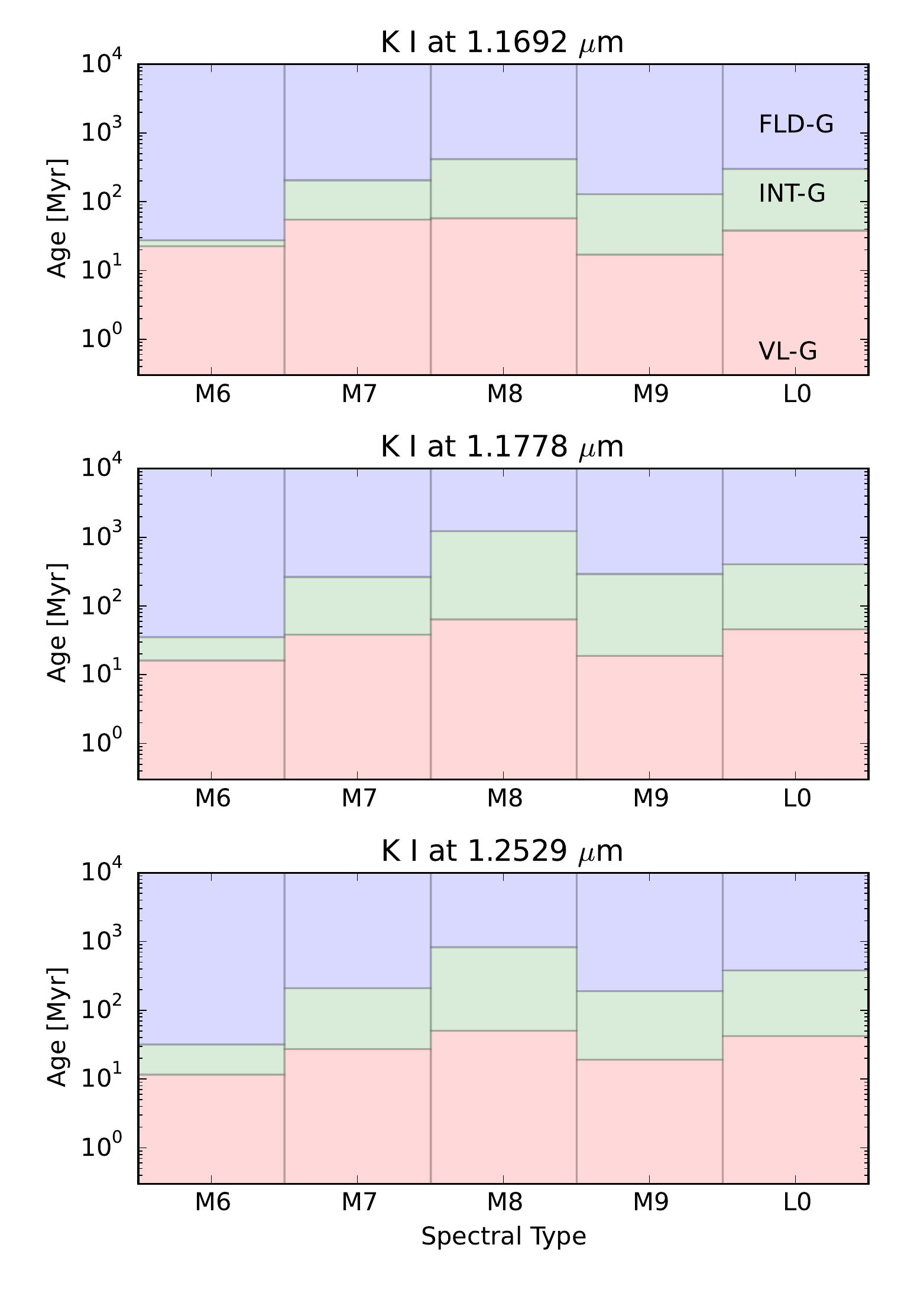}
\caption{Age vs. Spectral Type for each gravity index in A13. Ages were estimated by applying the best fit parameters for each spectral type bin and \kii \ line in Table~\ref{tab:ew_age} to the A13 EW boundaries between each gravity classification. Red shaded regions represent VL-G classifications, green shading denotes INT-G classification, and blue shading represents FLD-G ages. 
\label{fig:age_limits}}
\end{center}
\end{figure}

Although it is tempting to assign ages to each of the targets based on the relationships shown in Figure~\ref{fig:age_vs_ew} and Table~\ref{tab:ew_age}, we discourage against this because of the inability to significantly determine an age by combining the age estimates from each of the \kii \ EWs. Instead, we have determined broad age limits for each of the gravity classifications at each spectral type and for each \kii \ line. These are provided in Table~\ref{tab:age_ranges} and shown in Figure~\ref{fig:age_limits}. We determine the age at which each of the gravity classifications intersects the best-fit lines from Figure~\ref{fig:age_vs_ew} for each spectral type bin and each \kii \ line. The lower age limits are set by $\rho$ Oph at 0.3 Myr and the upper age limits are set by age estimates of the field population at $\sim$10 Gyr. The results in Table~\ref{tab:age_ranges} and Figure~\ref{fig:age_limits} show how large of a spread in age estimates there can be for objects of similar spectral types with varying gravity types. For example, an M8 target with an INT-G classification would be estimated to have an age ranging from 50 Myr -- 1.21 Gyr depending on the \kii \ line used. An M9 INT-G target on the other hand, has age estimates ranging from 17 -- 290 Myr. Additionally, the bounds of the intermediate age range for a single \kii \ index across all 5 spectral types show large variation. The most extreme example of the range in age estimates for a single \kii \ line is seen for the upper age limit probed by INT-G for the \kii \ line at 1.1778 \microm , which ranges from 35 Myr to 1.21 Gyr depending on spectral type. We find that assigning a specific age range to the A13 gravity classifications is beyond the scope of this paper. 

Uncertainty in age determination and gravity classification was also discussed in \citet{aller2016}. The authors of that paper determined uncertainties on the gravity types using a Monte Carlo approach in order to distinguish borderline objects that might otherwise be classified as FLD-G objects, but show hints of youth. The broad range of ages associated with each gravity classification indicates the importance of further age analysis using other techniques. For example, kinematic information, potential young moving group membership, or stellar/sub-stellar benchmark companions could further distinguish the age of a particular object. Low-mass stars and brown dwarfs exhibiting signs of low gravity merit follow-up observations to confirm or refute their potential youth status.

\subsection{The Future for Surface Gravity Analysis}

With the aforementioned limitations of gravity classification as a method for establishing the age of a young brown dwarf, we can make significant advances in this field in several ways. First, a larger sample of young L and T dwarfs of known ages is needed to extend the A13 classifications to later spectral types. In this paper, we do not have a large enough sample of young objects with spectral types later than $\sim$ L6 in order to expand these classifications into the L/T transition. The unshaded regions of Figures~\ref{fig:EW_SPT_MLT} and~\ref{fig:FeH_MLT} display the regime in which significant progress can be made in furthering our understanding of gravity and age at varying masses and temperatures for these substellar objects. 

Several recent works have highlighted discoveries in this area. Examples include the new bonafide T5.5 member of AB Doradus \citep{gagne2015c}, the young L7 TW Hya interloper presented in \citet{gagne2016}, the two new candidate L7 members of TW Hya presented in \citet{kellogg2016} and \citet{schneider2016}, and the 10 candidate YMG members of spectral type L7-T4.5 found with Pan-STARRS and WISE in \citet{best2015}. Additionally, discoveries of jovian exoplanets around young stars present a method for studying objects that appear very similar to young field L and T dwarfs. Some examples of these exciting discoveries include 51 Eri b \citep{macintosh2015} and GU Psc b \citep{naud2014}. Progress in extending the sample of known-age late-M and early-L dwarfs will also further our understanding of observational signatures of brown dwarf evolution. To this end, \citet{burgasser2016a} presented the first planetary-mass member of 32 Ori (L1). Kinematic information of M and L dwarfs with Gaia \citep{gaia2016a, gaia2016b} will confirm or refute the membership of young moving group candidates and allow for discoveries of new members. 

Second, high-resolution spectroscopy obtained with the next generation of 30-m class telescopes in conjunction with improved atmospheric models will allow us to better correlate surface gravity with age in these young brown dwarfs. The current atmospheric models have incomplete line lists and do not accurately represent the observed behavior of the \kii \ lines. The diversity of spectral features present in low-mass stars and brown dwarfs likely stems from physical properties and atmospheric conditions that we cannot probe at these moderate resolving powers or using these particular diagnostics. If atmospheric models continue to improve in tandem with observational capabilities, it may be possible to better isolate the effect that surface gravity has on brown dwarf spectral features.


\section{Summary}
We presented 228 $J$-band spectra of M, L, and T dwarfs in the BDSS, the largest set of publicly available NIR spectra at R $\sim$ 2000. Using the same J-band gravity sensitive indices as \citet{allers2013}, we calculated \kii \ equivalent widths and FeH absorption to determine gravity classifications for objects of spectral type M6--L7. Our technique is verified with 20 overlapping targets from A13, for which we derive similar gravity classifications despite using fewer spectral indicators. A subset of 73 objects with known (or suspected) ages from the literature (after excluding binaries from the full sample of known-age objects) define the trend of \kii \ EW with age. By assigning ages to the boundaries of each gravity designation for spectral types M6--L0, we find that the age ranges probed by each of the \kii \ lines vary widely. With a larger sample of age-calibrated M, L, and T dwarfs it will be possible to estimate ages for the entire sample with much greater certainty. This level of precision will likely require high signal-to-noise, high-resolution spectra of benchmark systems and detailed model comparison. Until then, the gravity designations from A13 remain a useful tool for dividing the low-mass products of star formation by relative age.


\section{Acknowledgements}
The authors wish to recognize and acknowledge the very significant cultural role and reverence that the summit of Mauna Kea has always had within the indigenous Hawaiian community.  We are most fortunate to have the opportunity to conduct observations from this mountain. This research has benefitted from the M, L, T, and Y dwarf compendium housed at DwarfArchives.org. This research has made use of the SIMBAD database, operated at CDS, Strasbourg, France. We thank the anonymous referee for their insightful comments, which improved the paper. 

Facilities: \facility{Keck:II(NIRSPEC)}


\bibliographystyle{apj}
\bibliography{bibidy}{}
\mbox{~}
\clearpage

\LongTables 
\begin{landscape} 
\center 
\begin{deluxetable}{lccccccccc} 
\tabletypesize{\scriptsize} 
\tablewidth{0in} 
\tablecaption{Observations and Designations \label{tab:obs}} 
\tablehead{ 
\colhead{Designation} & 
\colhead{Disc.} & 
\colhead{Short Name} & 
\colhead{SpT} & 
\colhead{SpT} & 
 \colhead{UT Date of} & 
\colhead{A0V} & 
\colhead{Slit Width} & 
\colhead{Exp. Time} \\ 
\colhead{} & 
\colhead{Ref} & 
\colhead{} & 
\colhead{} & 
\colhead{Ref} & 
\colhead{Observation} & 
\colhead{Calibrator} & 
\colhead{[arcsec]} & 
\colhead{[s]}
} 
\startdata 
2MASS J06195260-2903592&29&2MASS 0619-2903&M5&1&11/21/2004&HD 41473&0.38&1200\\ 
2MASS J00433134+2054316&68&LHS 1135&d/sdM5&54&12/4/2003&HD 7215&0.38&600\\ 
GJ 577 BC &64&Gl 577 BC&M5.5+M5.5&88&3/24/2003&HD 132072&0.38&600\\ 
2MASS J16262152-2426009&40&$\rho$ Oph GY 5&M5.5&81&5/14/2003&HD 151736&0.38&600\\ 
WDS J04325+1732Ba&27&GG Tau Ba&M6&103&11/7/2004&HD 28354&0.38&1200\\ 
2MASS J14594626+0004427&54&2MASS 1459+0004&M6&54&7/19/2005&HD 123233&0.38&1200\\ 
2MASS J22344161+4041387&29&2MASS 2234+4041AB&M6+M6&2&12/4/2003&BD +39 4890&0.38&600\\ 
2MASS J16051403-2406524&78&DENIS-P 1605-2406&M6&78&5/2/2004&HD 147384&0.38&1200\\
\enddata 
\end{deluxetable} 
\clearpage 
\end{landscape} 

\LongTables 
\begin{landscape} 
\center 
\begin{deluxetable}{llccccccccc} 
\tabletypesize{\tiny} 
\tablewidth{0pt} 

\tablecaption{Equivalent Widths and Spectral Index values for the BDSS J band spectra \label{tab:k1ew}} 
\tablehead{ 
\colhead{Name} & 
\colhead{SpT} & 
\colhead{K I EW ($\AA$)} & 
\colhead{K I EW ($\AA$)} & 
\colhead{K I EW ($\AA$)} & 
\colhead{K I EW ($\AA$)} & 
\colhead{FeH$_J$} & 
\colhead{Gravity Scores} & 
\colhead{Gravity Type} \\ 
\colhead{} & 
\colhead{} & 
\colhead{$1.1692 \mu m$} & 
\colhead{$1.1778 \mu m$} & 
\colhead{$1.2437 \mu m$} & 
\colhead{$1.2529 \mu m$} & 
\colhead{$1.20 \mu m$} & 
\colhead{K1 K2 K4 FeH} & 
\colhead{} 
 } 
\startdata 
2MASS 0619-2903&M5&0.1$\pm$0.64&0.52$\pm$0.58&0.41$\pm$1.24&1.51$\pm$0.57&1.021$\pm$0.014& n n 0. n &N/A\\ 
LHS 1135&d/sdM5&2.15$\pm$0.37&4.0$\pm$0.3&1.91$\pm$0.4&2.51$\pm$0.16&1.045$\pm$0.004&N/A &N/A \\ 
Gl 577 BC&M5.5&0.6$\pm$0.04&1.81$\pm$0.04&1.15$\pm$0.1&1.44$\pm$0.05&1.042$\pm$0.001&n n 0. n &N/A\\ 
Rho Oph GY 5&M5.5&0.55$\pm$0.82&0.5$\pm$0.46&0.81$\pm$0.61&0.65$\pm$0.24&1.021$\pm$0.01&n n 2. n &N/A\\ 
2MASS 1459+0004&M6&0.79$\pm$0.76&1.49$\pm$0.7&1.53$\pm$1.47&1.57$\pm$0.68&1.021$\pm$0.018&  2.  2.  2.  2.&VL-G\\ 
2MASS 2234+4041&M6&-0.02$\pm$0.64&0.53$\pm$0.51&0.97$\pm$0.83&0.15$\pm$0.3&1.026$\pm$0.011&  2.  2.  2.  2.&VL-G\\ 
DENIS-P 1605-2406&M6&0.1$\pm$0.1&0.74$\pm$0.09&0.8$\pm$0.2&0.53$\pm$0.09&1.039$\pm$0.002&  2.  2.  2.  2.&VL-G\\ 
KPNO Tau 3&M6&-0.71$\pm$0.17&-0.81$\pm$0.1&1.19$\pm$0.19&0.7$\pm$0.09&1.025$\pm$0.002&  2.  2.  2.  2.&VL-G\\
\enddata 
\end{deluxetable} 
\clearpage 
\end{landscape}

\LongTables
\begin{landscape} 
\center 
\begin{deluxetable}{lccccccccc} 
\tabletypesize{\small} 
\tablewidth{0in} 
\tablecaption{BDSS Objects With Known Ages \label{tab:ages}} 
\tablehead{ 
\colhead{Designation} &
\colhead{Short Name} &
\colhead{Right Ascension} & 
\colhead{Declination} &
\colhead{SpT} & 
\colhead{Age} & 
\colhead{Gravity Type} & 
\colhead{Notes} & 
\colhead{\% } & 
\colhead{Membership} \\
\colhead{} &
\colhead{} &
\colhead{} &
\colhead{} &
\colhead{} &
\colhead{[Myr]} &
\colhead{} &
\colhead{} &
\colhead{Membership} &
\colhead{Ref.} 
 } 
\startdata
2MASS J22443167+2043433&2MASS 2244+2043&22 44 31.674&+20 43 43.30&L6&$149\substack{+51 \\ -19}$&VL-G&AB Dor&99.6&7\\ 
\\ \hline \\
2MASS J03230483+4816111&AP 310&03 23 04.83&+48 16 11.2&M6&$90\pm10$&FLD-G&Alpha Persei&100&25\\ 
Cl* Melotte 20 AP 316& AP 316&03 27 01.3&+49 14 40&M6&$90\pm10$&FLD-G&Alpha Persei&100&25\\ 
2MASS J03180906+4925189&AP 301&03 18 09.06&+49 25 19.0&M6.5&$90\pm10$&FLD-G&Alpha Persei&100&25\\ 
2MASS J03204346+5059396&AP 270&03 20 43.47&+50 59 39.6&M7&$90\pm10$&VL-G&Alpha Persei&54&12\\ 
2MASS J03354735+4917430&AP 325&03 35 47.36&+49 17 43.1&M7&$90\pm10$&INT-G&Alpha Persei&62&12\\
\enddata 
\end{deluxetable} 
\clearpage
\end{landscape} 

\clearpage

\clearpage 
\begin{deluxetable}{lccccc} 
\tabletypesize{\small} 
\tablewidth{0in} 
\tablecaption{Best Fit Parameters for Age vs. \kii \ Equivalent Width \label{tab:ew_age} }
\tablehead{ 
\colhead{SpT} &
\colhead{A} & 
\colhead{3$\sigma_{A}$} & 
\colhead{B} & 
\colhead{3$\sigma_{B}$} \\
\colhead{} &
\colhead{[Myr]} & 
\colhead{} & 
\colhead{[$\AA^{-1}$]} & 
\colhead{}
 } 
\startdata
 & & \kii \ at 1.1692 \microm & & \\ \hline \\
M6$\pm$1& 1.21397 & 0.18509 & 0.82751 & 0.03330 \\ 
M7$\pm$1& 1.11259 & 0.13018 & 0.80449 & 0.02633 \\ 
M8$\pm$1& 0.96102 & 0.12268 & 0.66583 & 0.02163 \\ 
M9$\pm$1& 0.49177 & 0.08032 & 0.47582 & 0.01421 \\ 
L0$\pm$1& 1.51591 & 0.29961 & 0.37076 & 0.01813 \\

\\ \hline \\
& &  \kii \ at 1.1778 \microm & & \\ \hline \\
M6$\pm$1& 0.64072 & 0.09984 & 0.55605 & 0.02238 \\ 
M7$\pm$1& 0.49738 & 0.06485 & 0.58026 & 0.01879 \\ 
M8$\pm$1& 0.31632 & 0.04927 & 0.59087 & 0.01819 \\ 
M9$\pm$1& 0.23653 & 0.04466 & 0.42362 & 0.01261 \\ 
L0$\pm$1& 1.85884 & 0.29051 & 0.27824 & 0.01156 \\

\\ \hline \\
 & & \kii \ at 1.2529 \microm & & \\ \hline \\
M6$\pm$1& 0.43282 & 0.07065 & 0.84337 & 0.03082 \\ 
M7$\pm$1& 0.46584 & 0.06145 & 0.80843 & 0.02697 \\ 
M8$\pm$1& 0.46064 & 0.06628 & 0.77101 & 0.02500 \\ 
M9$\pm$1& 0.59026 & 0.07495 & 0.49215 & 0.01194 \\ 
L0$\pm$1& 1.86551 & 0.29311 & 0.38854 & 0.01493 \\

\tablecomments{The parameters in this table may be applied to determine an age estimate using the following equation: Age[Myr] = $A \times 10^{B  \times EW [\AA]}$}
\enddata 
\end{deluxetable} 
\begin{deluxetable}{c@{\hskip .8cm}ccc@{\hskip .6cm}ccc@{\hskip .6cm}lcr} 
\tabletypesize{\small} 
\tablewidth{0in} 
\tablecaption{Age Ranges For A13 Gravity Classifications for \kii \ EWs \label{tab:age_ranges} }
\tablehead{ 
\colhead{} & 
\multicolumn{9}{c}{Age Ranges [Myr]} \\ [5pt] \hline \\ 
\colhead{SpT} &
\multicolumn{3}{c}{\kii \ at 1.1692 \microm } & 
\multicolumn{3}{c}{\kii \ at 1.1778 \microm } &
\multicolumn{3}{c}{\kii \ at 1.2529 \microm}  \\ [5pt] 
\colhead{} &
\colhead{VL-G} & 
\colhead{INT-G} & 
\colhead{FLD-G} & 
\colhead{VL-G} &
\colhead{INT-G} &
\colhead{FLD-G} & 
\colhead{VL-G} & 
\colhead{INT-G} & 
\colhead{FLD-G}
 } 
\startdata

\hspace{.3cm}M6&0.3--22&22--27&27--10 Gyr &0.3--16&16--35&35--10 Gyr&0.3--12&12--32&32--10 Gyr \\ [5pt]
\hspace{.3cm}M7&0.3--55&55--205&205--10 Gyr&0.3--39&39--262&262--10 Gyr&0.3--27&27--208&208--10 Gyr \\ [5pt]
\hspace{.3cm}M8&0.3--57&57--413&413--10 Gyr&0.3--64&64--1.21 Gyr&1.21 Gyr--10 Gyr&0.3--50&50--817&817--10 Gyr \\ [5pt]
\hspace{.3cm}M9&0.3--17&17--128&128--10 Gyr&0.3--19&19--290&290--10 Gyr&0.3--19&19--190&190--10 Gyr \\ [5pt]
\hspace{.3cm}L0&0.3--38&38--296&296--10 Gyr&0.3--46&46--403&403--10 Gyr&0.3--42&42--378&378--10 Gyr

\enddata 
\end{deluxetable} 

\clearpage

\begin{figure}[H]
\begin{center}
\makebox[\textwidth][c]{\includegraphics[scale=.8,angle=0]{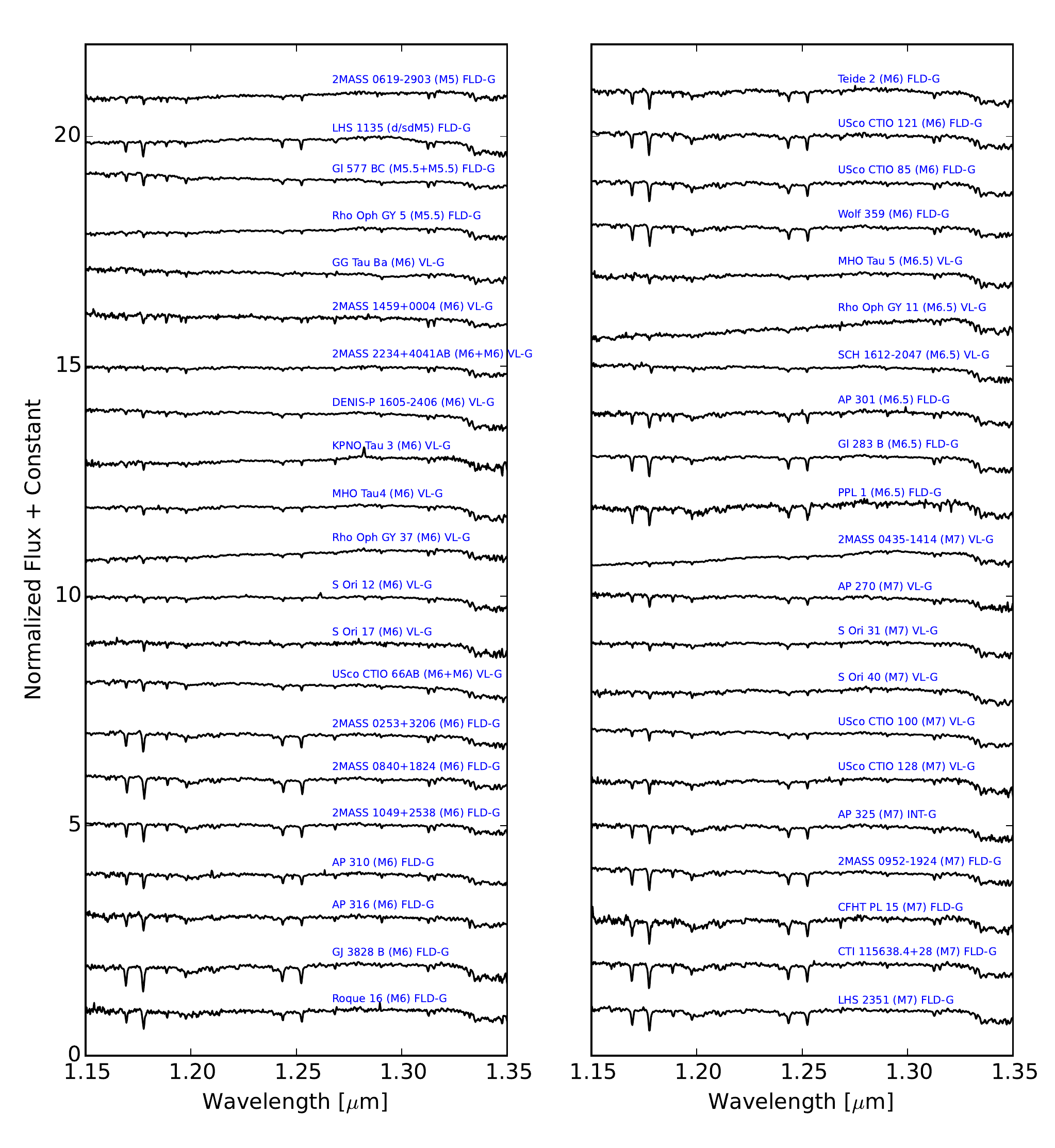}}
\caption{M5--M7 dwarfs, ordered by spectral type and then surface gravity type (if applicable)
\label{fig:appendix1}}
\end{center}
\end{figure}

\begin{figure}
\begin{center}
\makebox[\textwidth][c]{\includegraphics[height=9in, width= 8in, angle=0]{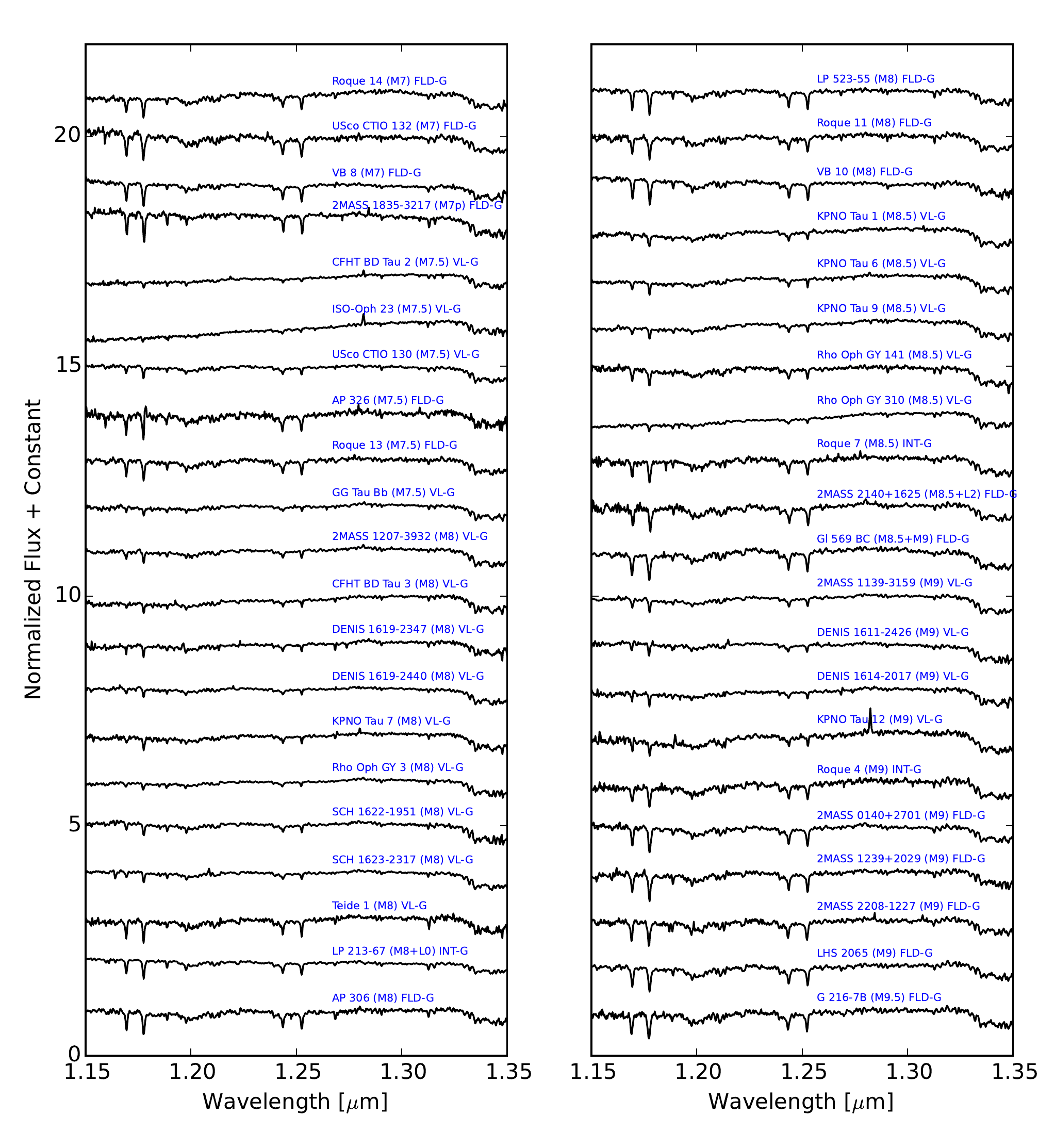}}
\caption{M7--M9.5 dwarfs, ordered by spectral type and then surface gravity type (if applicable)
\label{fig:appendix2}}
\end{center}
\end{figure}

\begin{figure}
\begin{center}
\makebox[\textwidth][c]{\includegraphics[height=9in, width= 8in,angle=0]{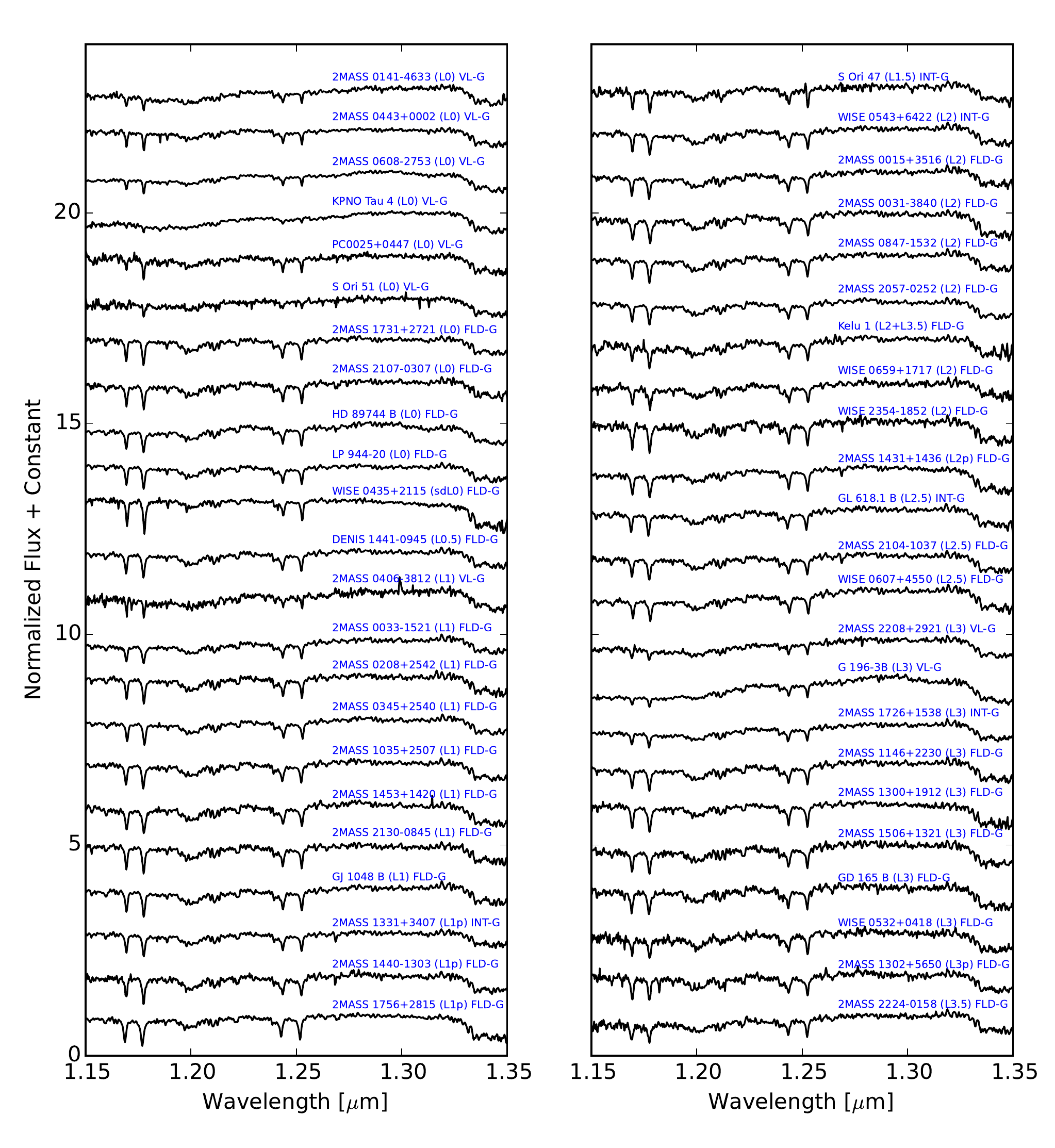}}
\caption{L0--L3.5 dwarfs, ordered by spectral type and then surface gravity type (if applicable)
\label{fig:appendix3}}
\end{center}
\end{figure}

\begin{figure}
\begin{center}
\makebox[\textwidth][c]{\includegraphics[height=9in, width= 8in,angle=0]{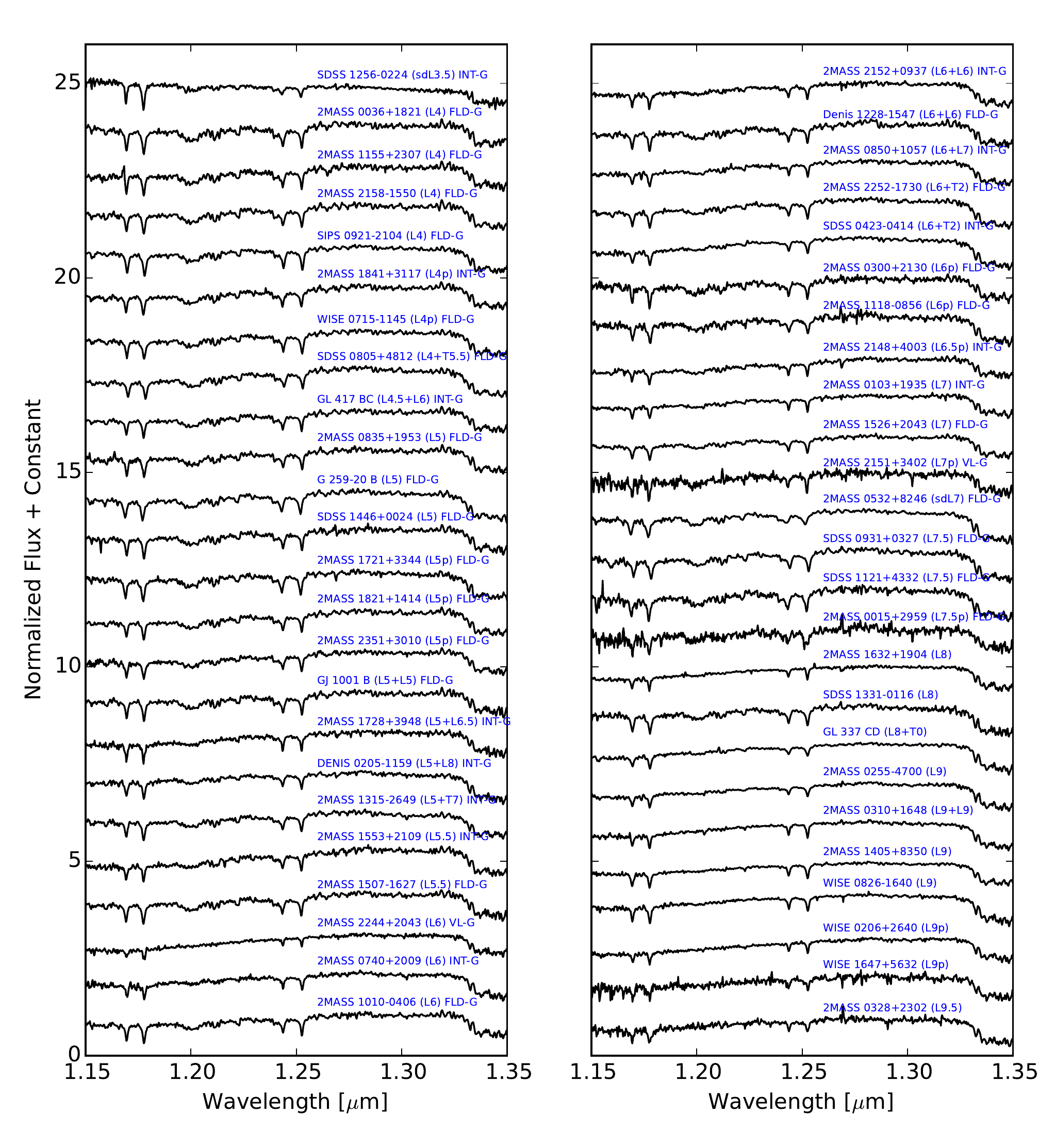}}
\caption{L3.5--L9.5 dwarfs, ordered by spectral type and then surface gravity type (if applicable)
\label{fig:appendix4}}
\end{center}
\end{figure}

\begin{figure}
\begin{center}
\makebox[\textwidth][c]{\includegraphics[height=9in, width= 8in,angle=0]{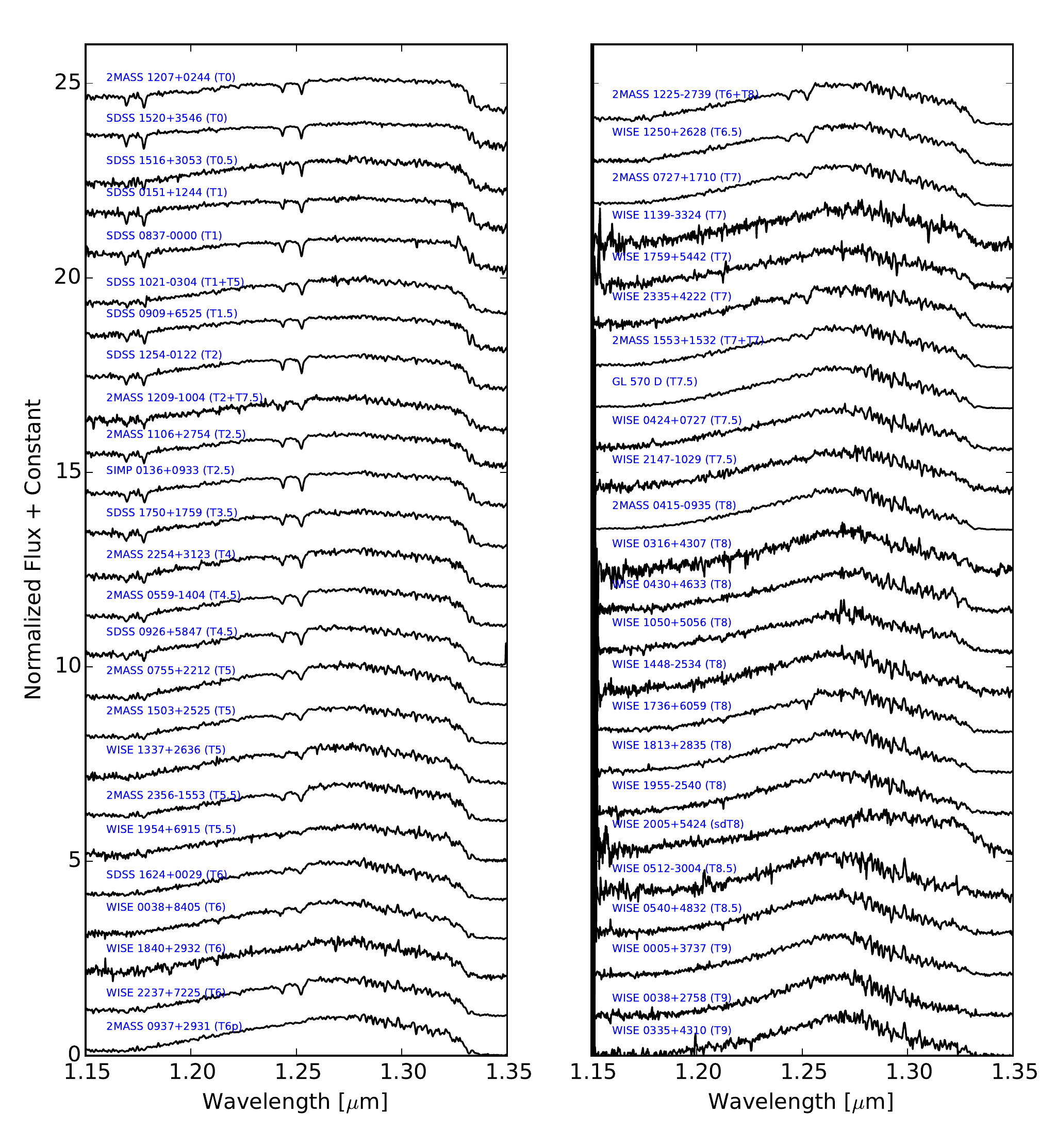}}
\caption{T0--T9 dwarfs, ordered by spectral type and then surface gravity type (if applicable)
\label{fig:appendix5}}
\end{center}
\end{figure}

\begin{appendix}
Here we present all $J$-band spectra for the BDSS, ordered by spectral type and then surface gravity (if applicable) in Figures~\ref{fig:appendix1},~\ref{fig:appendix2},~\ref{fig:appendix3},~\ref{fig:appendix4}, and ~\ref{fig:appendix5}. Many of these spectra were previously published in \citet{mclean2003}, \citet{mcgovern2004}, \citet{kirkpatrick2010}, \citet{mace2013a}, but the majority are published here for the first time. All spectra will be available for download on bdssarchive.org. In addition to the \kii , FeH, VO, and \wat \ absorption features noted in Figure~\ref{fig:L3_example}, some spectra have the Al doublet at 1.32 \microm , the Pa$\beta$ emission line at 1.28 \microm , and some show significant reddening. The A13 indices were shown to be robust against reddening, so this should not affect our results.

\end{appendix}

\end{document}